\documentclass[aps, prd, preprint,,amsmath,amssymb，eshowkeys,nofootinbib]{revtex4-1}
\usepackage{verbatim,graphics,graphicx,color,slashed,textcomp,bbm,mathdots,multirow,array}
\usepackage{tikz}
\usepackage[normalem]{ulem} 
\usepackage[colorlinks=true,linkcolor=red,urlcolor=blue,citecolor=blue]{hyperref}
\usepackage{subfigure}
\usepackage{tabularx}

\usepackage{autobreak}
\usepackage{float}

\graphicspath{{fig/}}

\makeatletter

\newcommand{\Rmnum}[1]{\expandafter\@slowromancap\romannumeral #1@}
\makeatother

\usepackage[colorlinks=true,linkcolor=red,urlcolor=blue,citecolor=blue]{hyperref}

\begin{document}

\title{Distinguishing Monochromatic Signals in LISA and Taiji: Ultralight Dark Matter versus Gravitational Waves}
\author{Heng-Tao Xu$^{a,b}$}
\author{Yue-Hui Yao$^{c,d}$}
\author{Yong Tang$^{a,c,d}$}
\author{Yue-Liang Wu$^{a,b,c,d}$}

\affiliation{\begin{footnotesize}
		${}^a$School of Fundamental Physics and Mathematical Sciences, Hangzhou Institute for Advanced Study, UCAS, Hangzhou 310024, China \\        
            ${}^b$Institute of Theoretical Physics, Chinese Academy of Sciences, Beijing 100190, China \\
		${}^c$University of Chinese Academy of Sciences (UCAS), Beijing 100049, China\\
        ${}^d$International Center for Theoretical Physics Asia-Pacific, Beijing 100190, China\\
		\end{footnotesize}}

\begin{abstract}
Ultralight dark matter (ULDM) is an attractive candidate for cold dark matter, one of the main mysterious components of the Universe.
Recent studies suggest that gravitational-wave (GW) laser interferometers can also 
detect bosonic ULDM fields, which would produce monochromatic signals resembling those from gravitational waves~(GWs). Distinguishing between these potential origins therefore would be essential.
In this work, we develop a method to address this challenge for space-based GW interferometers (such as LISA and Taiji) by utilizing the null-response channel (NRC) in interferometric combinations, a channel constructed to have zero response to a specific type of source from a given direction. 
We find that while the GW NRC remains blind to GWs from a specific direction, it still responds to ULDM, particularly at frequencies above the interferometer's critical frequency. The ULDM NRC exhibits similar behavior. 
Based on these observations, 
we outline a test procedure to discriminate between signal origins. Our method provides a new diagnostic tool for analyzing monochromatic signals in space-based GW interferometers, potentially expanding the scientific scope of future missions.
\end{abstract}
 
\maketitle

\section{Introduction}
The first detection of gravitational waves~(GWs)~\cite{LIGOScientific:2016aoc} has marked the dawn of gravitational-wave (GW) astronomy. With the advent of next-generation detectors, both ground-based~\cite{Punturo:2010zz, Reitze:2019iox} and space-based~\cite{amaroseoane2017laser, Hu:2017mde, Luo_2016}, GWs would powerfully help to explore the dark side of our Universe. While GW interferometers are primarily designed to detect GWs, they are also highly sensitive to other physical phenomena~\cite{Pierce:2018xmy, Morisaki:2018htj, Grote:2019uvn, Guo:2019ker, Vermeulen:2021epa, Fukusumi:2023kqd, Kim:2023pkx, PhysRevD.110.023025, Yu:2023iog, Yao:2024fie, Gue:2024txz, Yao:2024hap, Yao:2025wfd, Zhang:2025fck, PhysRevD.109.095012}, which may also be probed via the associated GWs~\cite{PhysRevD.110.115026, PhysRevD.111.103511, Su:2025nkl} or through their influence on GW signals~\cite{Blas:2024duy, Nomura:2024cku, Chen:2025jch, Feng:2025fkc}. 

For instance, many theories beyond the standard model~(SM) of particle physics and cosmology predict the existence of light bosonic fields~\cite{PhysRevLett.38.1440,Weinberg1978,PhysRevLett.40.279,PRESKILL1983127,Dine:1982ah,Damour:1994ya,Damour:1994zq,Capozziello:2011et,PhysRevD.84.103501,PhysRevD.93.103520,Ema:2019yrd}, which are viable dark matter~(DM) candidates~\cite{Cirelli:2024ssz, Hui:2021tkt, Ferreira:2020fam} and can interact with SM particles very weakly.
In the presence of such a DM field within the solar system, its oscillations can exert external forces on the test masses, inducing motion that leads to variation in the distance and thus produce detectable signals. In most cases, signals induced by such ultralight DM (ULDM) are quasi-monochromatic, with frequencies at integer multiples of the ULDM's Compton frequency~\cite{Morisaki:2018htj, Guo:2019ker, Kim:2023pkx, PhysRevD.107.063015, Amaral:2024tjg, Gue:2024txz}.
However, quasi-monochromatic signals can also originate from GWs, such as those from compact binaries inspirals~\cite{Sathyaprakash:2009xs}.
Therefore, when such signals are detected, it is crucial to determine whether they originate from GWs or DM.

In this work, we address this question in the context of space-based GW interferometers, such as LISA~\cite{amaroseoane2017laser} and Taiji~\cite{Hu:2017mde}. Previously we have suggested~\cite{Yu:2023iog, Yao:2024fie} that one can differentiate between a GW signal and a ULDM one by exploiting the differences in the responses of interferometry channels, where the argument relies on sky-averaged responses and thus neglects the direction dependence of the responses. Here, we leverage the directional dependence and present a new and more systematic approach. 
We propose a null-response channel (NRC) for ULDM and compare it with the case~\cite{Tinto:2004nz,Barroso:2024azl} for GWs.
We systematically study the response of the GW NRC to ULDM and vice versa.
We find that, while the GW NRC remains blind to GWs throughout the band, its response to ULDM differs below and above the interferometer’s critical frequency.
Below this frequency, the ULDM signal is strongly suppressed, whereas above it, the GW NRC responds effectively to ULDM.
Similar behavior is observed for the ULDM NRC to GWs.
This allows for a test procedure to distinguish signal origins, as summarized in Fig.~\ref{fig:exam}. Our method provides a new diagnostic tool for the monochromatic signals in space-based GW interferometers.

This paper is organized as follows. 
In Sec.~\ref{sec:ZS}, we first introduce the idea behind the NRC and generalize it into a form facilitating our discussion of ULDM.
We then construct the NRCs for vector and scalar ULDM.
In Sec.~\ref{sec:apply}, we examine the responses of both GW NRC and ULDM NRC and outline the procedure for identifying the origin of a detected monochromatic signal in LISA and Taiji.
In Sec.~\ref{sec:limit}, we discuss the validity of our method and possible further investigations on more practical cases.
Finally, we conclude in Sec.~\ref{Conclusions}.

Throughout the paper we use natural units ($c=\hbar=1$) and refer to ULDM and bosonic field/wave interchangeably.

\section{Theoretical framework} \label{sec:ZS}
\subsection{Detector and Signal Response} \label{sec:corona}
\begin{figure}[t]
    \centering
    \includegraphics[width=0.9\linewidth]{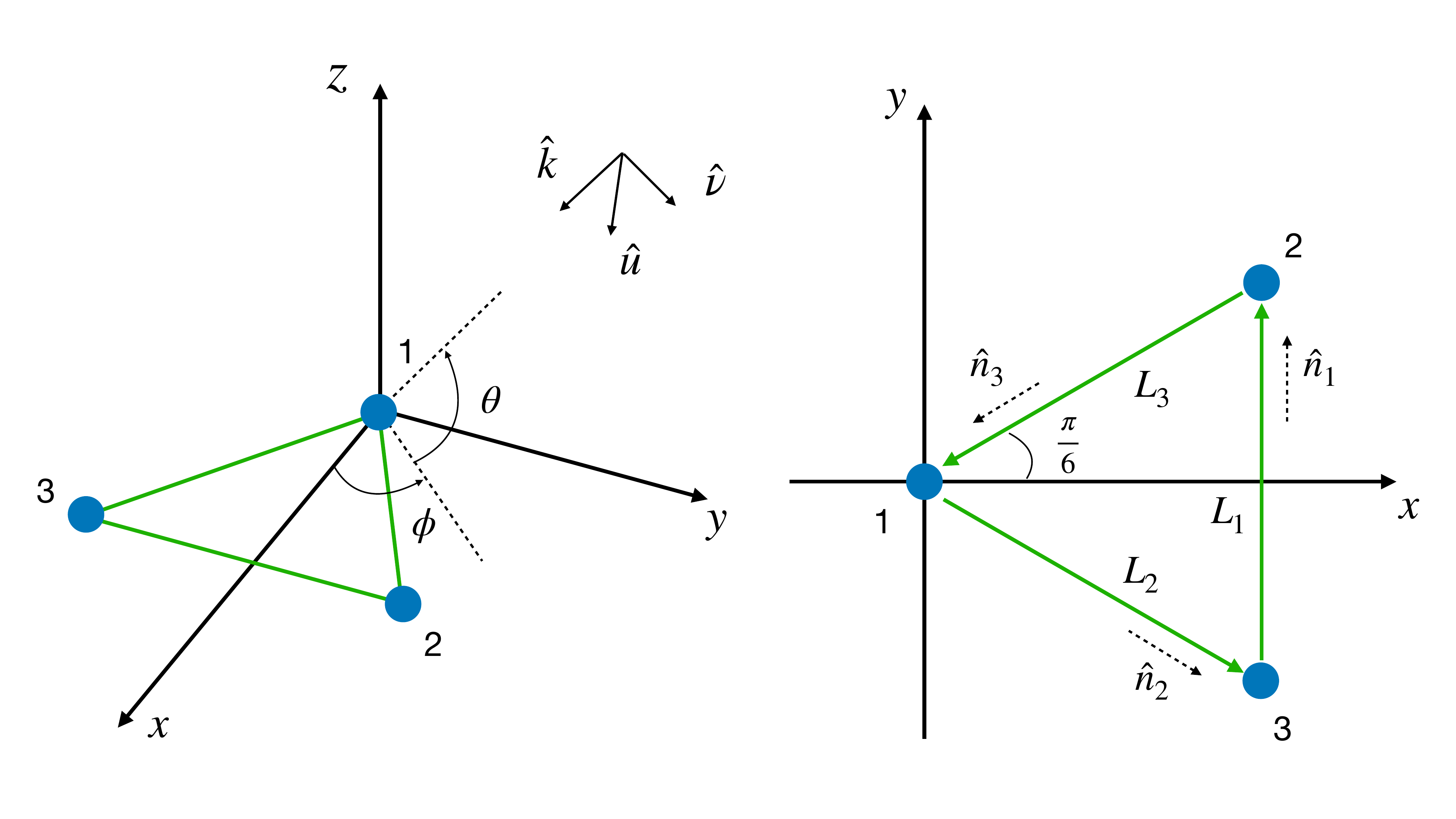}
    \caption{Schematic of the detector and coordinate system. The three spacecraft are located at $\mathbf{x}_1 = (0,0,0)$, $\mathbf{x}_2 = L(\cos\frac{\pi}{6},\sin\frac{\pi}{6},0)$, $\mathbf{x}_3 = L(\cos\frac{\pi}{6},-\sin\frac{\pi}{6},0)$, 
    where $L$ is the arm length of the detector.
    The unit vectors along the arms are given by $\hat{n}_1=\frac{1}{L}(\mathbf{x}_2-\mathbf{x}_3),\hat{n}_2=\frac{1}{L}(\mathbf{x}_3-\mathbf{x}_1),\hat{n}_3=\frac{1}{L}(\mathbf{x}_1-\mathbf{x}_2)$. 
    The polar angle $\theta$ is measured from the detector plane and ranges from $-\pi/2$ to $\pi/2$, with $\theta=\pi/2$ corresponding to the positive $z$-axis.}
    \label{fig:cor}
\end{figure}

A space-based GW laser interferometer~\cite{amaroseoane2017laser, Hu:2017mde} typically consists of three spacecrafts arranged in a quasi-equilateral triangle constellation, as illustrated in Fig.~\ref{fig:cor}. Each spacecraft hosts two optical benches and two free-falling test masses. Laser beams are exchanged between spacecrafts to monitor the inter-spacecraft distances. Each spacecraft both sends and receives laser beams to and from the other two, forming six laser links.

The basic observables are the six single-link data streams, which record
the relative frequency fluctuations of laser:
\begin{equation}
    y_{ij}(t) = y^s_{ij}(t) + y^n_{ij}(t),
\end{equation}
where $ i,j=1,2,3$ and $i\neq j$. The symbols $y^s_{ij}(t)$ and $y^n_{ij}(t)$ represent the collective contributions of signals and noises, respectively.
For a GW propagating in the direction $\hat{k}$, the signal is given by~\cite{Babak:2021mhe}
\begin{equation} \label{eq:1link gw}
    y^{\text{gw}}_{ij}(t) = - \sum_{p=+,\times}\frac{\hat{n}_{ij}\otimes \hat{n}_{ij}:\mathbf{e}^{p}}{2(1-\hat{k}\cdot \hat{n}_{ij})}[h_p(t-\hat{k}\cdot \mathbf{x}_i) - h_p(t-L_{ij}-\hat{k}\cdot \mathbf{x}_j)],
\end{equation}
where $\mathbf{x}_i$ and $\mathbf{x}_j$ are the position vectors of receiving and sending spacecraft, respectively,
$\hat{n}_{ij} = (\mathbf{x}_i-\mathbf{x}_j)/L_{ij}$, $L_{ij} = |\mathbf{x}_i-\mathbf{x}_j|$,
$h_p(t)$ are the time-domain waveforms, and $\mathbf{e}^p$ are the polarization tensors.

We consider the scenario in which ULDM interacts directly with the SM particles.
In the presence of ULDM, the test masses, which are made of SM particles, experience external forces exerted by ULDM. The magnitude of these forces depends on local density of ULDM or the field value at the location of each test mass.
As a result, test masses at both ends of a link experience different forces, leading to relative motion. 
Modeled as the monochromatic plane-wave form, ULDM induces the signal~\cite{Morisaki:2018htj, Pierce:2018xmy, Yu:2023iog}
\begin{equation} \label{eq:1link bf}
y^{\text{bf}}_{ij}(t) = -e^{i\omega t}\sum_p H_p(\hat{e}_p \cdot \hat{n}_{ij})
\left[e^{-i\omega v\hat{k}\cdot \mathbf{x}_i} - e^{-i\omega(L_{ij} + v\hat{k}\cdot \mathbf{x}_j)}\right], 
\end{equation}
where $\hat{e}_p$ are the unit polarization vectors and $H_p$ is related to the amplitude of ULDM. Details will be provided later in the discussion around Eq.~(\ref{eq:vec field}).

However, given the current laser stability, the single-link data streams are dominated by the laser noise in $y^n_{ij}(t)$. 
Moreover, due to the complex orbital motion of the spacecraft, the interferometer’s arm lengths vary with time and are generally unequal. As a result, the simple Michelson interferometric configuration fail to effectively cancel the laser noise. 
To overcome these issues, time-delay interferometry~(TDI) is employed~\cite{PhysRevD.62.042002, PhysRevD.65.102002, PhysRevD.72.042003, Tinto:2020fcc, Wang:2020pkk}.
By appropriately time-shifting and combining the single-link data streams,
TDI synthesizes virtual equal-arm interferometric configurations so that the laser noise is strongly suppressed. 

There are various interferometric configurations, also referred to as TDI combinations or channels.
For example, Michelson-like $X$ combination in the first-generation TDI is given by
\begin{equation}
    X(t) = \left(y_{13} + y_{31,2} + y_{12,22} + y_{21,322}\right)
    - \left(y_{12} + y_{21,3} + y_{13,33} + y_{31,233}\right),
\end{equation}
where $y_{ij,mn}(t) = y_{ij}(t - L_m - L_n)$ with $L_m$ the length of the arm opposite spacecraft $m$. 
The Sagnac $\alpha$ combination is given by
\begin{align} \label{eq:def sagnac}
    \alpha = (y_{13}+y_{32,2}+y_{21,12}) - (y_{12}+y_{23,3}+y_{31,13}).
\end{align}
With permutation $1\rightarrow2\rightarrow3\rightarrow1$ we can get $\beta$ and $\gamma$ combinations. Not all these combinations are independent. The space of all TDI combinations can be generated by the four Sagnac combinations $\alpha$, $\beta$, $\gamma$ and $\zeta$~\cite{Tinto:2020fcc}, i.e., any combination can be expressed as a linear combination of these in the Fourier domain:
\begin{equation} \label{eq:general eta}
    \tilde{\eta} = a_1\tilde{\alpha} + a_2\tilde{\beta} + a_3\tilde{\gamma} + a_4\tilde{\zeta},
\end{equation}
where $\tilde{\;}$ denotes the counterpart of a quantity in the Fourier space, and $a_i$ are the frequency-dependent coefficients.
Due to the algebraic structure associated with time delay operators, the four generators are not independent~\cite{Tinto:2020fcc}.
At most frequencies, $\tilde{\zeta}$ can be expressed in terms of the remaining three and Eq.~(\ref{eq:general eta}) reduces to 
\begin{equation}
    \tilde{\eta} = a_1\tilde{\alpha} + a_2\tilde{\beta} + a_3\tilde{\gamma}.
\end{equation}
The matched filtering signal-to-noise ratio~(SNR) in $\eta$
is given by~\cite{PhysRevD.66.122002}
\begin{equation} \label{eq:eta SNR}
    \text{SNR}^2_{\eta} = \int df\;\frac{\left|a_1\tilde{\alpha}_s + a_2\tilde{\beta}_s + a_3\tilde{\gamma}_s\right|^2}
    {\mathbf{a}^{\dagger}\mathbf{S}_{\alpha}(f)\mathbf{a}} ,
\end{equation}
where $\tilde{\alpha}_s$ denotes the signal contribution,
$\mathbf{a} = [a_1, a_2, a_3]^T$ is the coefficient vector,
and $\mathbf{S}_{\alpha}(f)$ is the noise matrix of Sagnac combinations, which is symmetric and positive-definite. The elements of $\mathbf{S}_{\alpha}(f)$ are provided in Appendix~\ref{ap:noise}. 

The null-response channel~(NRC), as suggested by its name, is defined as the combination yielding zero $\text{SNR}$ for sources in a specified direction.
Since the integrand in Eq.~(\ref{eq:eta SNR}) is positive-definite,
the condition is satisfied only when
\begin{equation} \label{eq:eta def}
    a_1\tilde{\alpha}_s + a_2\tilde{\beta}_s + a_3\tilde{\gamma}_s = 0 .
\end{equation}
For GWs in general relativity~(GR), $\tilde{\alpha}_s$ can be decomposed into contributions from the two polarization modes:
\begin{equation} \label{eq:alpha decp}
    \tilde{\alpha}_s =
    \alpha^{\text{gw}}_{+}(f,\hat{k})\tilde{h}_+(f) + \alpha^{\text{gw}}_{\times}(f,\hat{k})\tilde{h}_{\times}(f) ,
\end{equation}
where $\tilde{h}_p(f)$ is the waveform in the Fourier domain, $\hat{k}$ points from the source to the origin of the coordinate system.
The symbol $\alpha^{\text{gw}}_{p}(f,\hat{k})$ denotes the response of $\alpha$ combination to GWs.
The explicit expressions for the responses of Sagnac combinations are provided in Appendix~\ref{ap:lfl_s}.

Substituting Eq.~(\ref{eq:alpha decp}) into Eq.~(\ref{eq:eta def}), we have
\begin{equation} \label{eq:eta GW}
    \left(\mathbf{a}\cdot\mathbf{x}^{\text{gw}}_{+}(\hat{k})\right)\tilde{h}_+
    + \left(\mathbf{a}\cdot\mathbf{x}^{\text{gw}}_{\times}(\hat{k})\right)\tilde{h}_{\times}
    = 0,
\end{equation}
where $\mathbf{x}^{\text{gw}}_p(\hat{k}) = [\alpha^{\text{gw}}_p(\hat{k}), \beta^{\text{gw}}_p(\hat{k}), \gamma^{\text{gw}}_p(\hat{k})]^T$ are the response vectors.
We omit the frequency dependency here for notation simplicity and restore it when necessary.
In general, the two response vectors are not parallel to each other, and 
a non-trivial solution of Eq.~(\ref{eq:eta GW}) is given by
\begin{equation} \label{eq:NRC gw}
    \mathbf{a}^{\text{gw}}_{c}(\hat{k}) = \mathbf{x}^{\text{gw}}_{+}(\hat{k}) \times \mathbf{x}^{\text{gw}}_{\times}(\hat{k}) .
\end{equation}
Therefore, the GW NRC is given by
\begin{equation}
    \tilde{\eta}^{\text{gw}}_c(f,\hat{k};\hat{k}_c) = \mathbf{a}^{\text{gw}}_c(\hat{k}_c)\cdot
    [\tilde{\alpha},\tilde{\beta},\tilde{\gamma}] ,
\end{equation}
which has zero SNR for GWs propagating in the direction specified by the parameter $\hat{k}_c$.
Note that, by the construction in Eq.~(\ref{eq:NRC gw}), the effectiveness of NRC does not depend on the detailed waveform.

The expression in Eq.~(\ref{eq:eta GW}) is instructive
and can be generalized to scenarios involving $n$ independent combinations (detectors) and other exotic signals, such as those arising from additional polarization modes in theories beyond GR or from ULDM.
The signal in a general combination can be expressed as
\begin{equation} \label{eq:eta general}
    \tilde{\eta}_s = \sum_p \left(\mathbf{a}\cdot\mathbf{x}_{p}(\hat{k}_p)\right) \tilde{h}_p,
\end{equation}
where $\tilde{h}_p$ is the waveform of a wave traveling in $\hat{k}_p$ produced by source $p$.
The coefficient vector and response vector are given by $\mathbf{a} = [a_1, \cdots, a_{n}]^T$ and $\mathbf{x}_{p}(\hat{k}_p) = [x^1_{p}(\hat{k}_p), \cdots, x^n_{p}(\hat{k}_p)]^T$, respectively, where $x^i_{p}$ is the response of the $i$th combination to source $p$.
Therefore, as suggesting by Eq.~(\ref{eq:eta general}), finding a NRC is equivalent to determining a coefficient vector $\mathbf{a}$ that is orthogonal to $p$ independent vectors $\mathbf{x}_{p}$ in an $n$-dimensional vector space.

\subsection{The NRC for a vector ULDM} \label{sec:NRC_vf}
In this section, we construct the NRC for a nonrelativistic vector ULDM field.
We describe the vector ULDM field by a monochromatic plane wave\footnote{Since DM particles follow a velocity distribution, ULDM field should be a superposition of plane waves~\cite{PhysRevA.97.042506, PhysRevD.97.123006, PhysRevD.111.015028}. However, on a timescale shorter than the field's coherence time, the field can be modeled as a plane wave with a random amplitude and phase drawn from appropriate probability distributions~\cite{Khmelnitsky:2013lxt,Yao:2024fie}.}
\begin{equation} \label{eq:vec field}
    \mathbf{A}(x) = 
    \sum_{p=1}^3 A_p \hat{e}_p e^{i(\omega t - \mathbf{k}\cdot \mathbf{x})},
\end{equation}
where $A_p$ are the amplitudes, $\hat{e}_p$ the unit polarization vectors, $\omega =2\pi f \simeq m$, and $\mathbf{k} \simeq mv \hat{k}$, $v=10^{-3}$. 
Following the idea explained in Sec.~\ref{sec:corona}, a vector ULDM signal can be expressed as
\begin{equation} \label{eq:eta vf}
    \tilde{\eta}_s = \sum_{p=1}^3 \left(\mathbf{a} \cdot \mathbf{x}^{\text{vf}}_{p}(\hat{k})\right) H_p T,
\end{equation}
where $H_p = gA_p$ are the dimensionless effective amplitudes, with $g$ encapsulating coupling strength and experiment-specific parameters, and $T$ is the observation duration.
The response vector is given by
$\mathbf{x}^{\text{vf}}_{p}(\hat{k}) = [\alpha^{\text{vf}}_p(\hat{k}), \beta^{\text{vf}}_p(\hat{k}), \gamma^{\text{vf}}_p(\hat{k})]^T$, where $\alpha^{\text{vf}}_p(\hat{k})$ is the response of $\alpha$ combination to vector fields.
The explicit expressions for responses are provided in Appendix~\ref{ap:lfl_s}.

At first glance, the presence of three polarizations implies that finding the vector NRC would be impossible, as there are only three independent combinations.
However, this changes upon closer examination of the response.
We first consider the simple static case in which the three spacecrafts that form the detector constellation lie in a common plane, and define polarization modes relative to it. We shall discuss the nonstatic case in Sec.~\ref{sec:limit}.

Since the response is proportional to $\hat{n}_{ij}\cdot\hat{e}_p$, see Eq.~(\ref{eq:1link bf}),
the combinations are inherently insensitive to the polarization mode orthogonal to the plane,
and Eq.~(\ref{eq:eta vf}) reduces to
\begin{equation}
    \tilde{\eta}_s = \sum_{p=1}^2 \left(\mathbf{a}\cdot\mathbf{x}^{\text{vf}}_{p}(\hat{k})\right) H_pT,
\end{equation}
where $p=1,2$ refer to the two polarization modes parallel to the plane.
Therefore, analogous to the GW NRC, the NRC for vector fields is given by
\begin{equation} \label{eq:NRC_vf}
    \mathbf{a}^{\text{vf}}_{c}(\hat{k}_c) = \mathbf{x}^{\text{vf}}_{1}(\hat{k}_c) \times \mathbf{x}^{\text{vf}}_{2}(\hat{k}_c) ,
\end{equation}
where $\mathbf{x}^{\text{vf}}_{i}$ represent the response vectors to the two parallel polarization modes. Interestingly, as shown in Appendix~\ref{ap:coefficients}, there exists a connection between the vector NRC and the symmetric Sagnac combination $\zeta$. 

The above discussions do not depend on the coordinate system, so we can proceed in a specific coordinate system, as illustrated in Fig.~\ref{fig:cor}.
We assume equal arm lengths and adopt the parameters of Taiji, as detailed in the Appendix~\ref{ap:noise}.
In Fig.~\ref{fig:zs_vdm}, we simulate
a vector signal in the vector NRC, $\tilde{\eta}^{\text{vf}}_c(f,\hat{k};\hat{k}_c)$.
The signal is generated by a transverse, circularly-polarized vector plane wave propagating in the direction $\hat{k} = (\theta, \phi) = (0.80,1.77)$\footnote{The direction is given by $\hat{k}=(-\cos\theta \cos\phi,-\cos\theta \sin\phi,-\sin\theta)$. All angles are in unit of radian.} with $f=1~\text{mHz}$ and unit amplitude.
As shown in Fig.~\ref{fig:zs_vdm}, the signal is strongly suppressed when $\hat{k}_c = \hat{k}$.
\begin{figure}[t]
    \centering
    \includegraphics[width=0.60\linewidth]{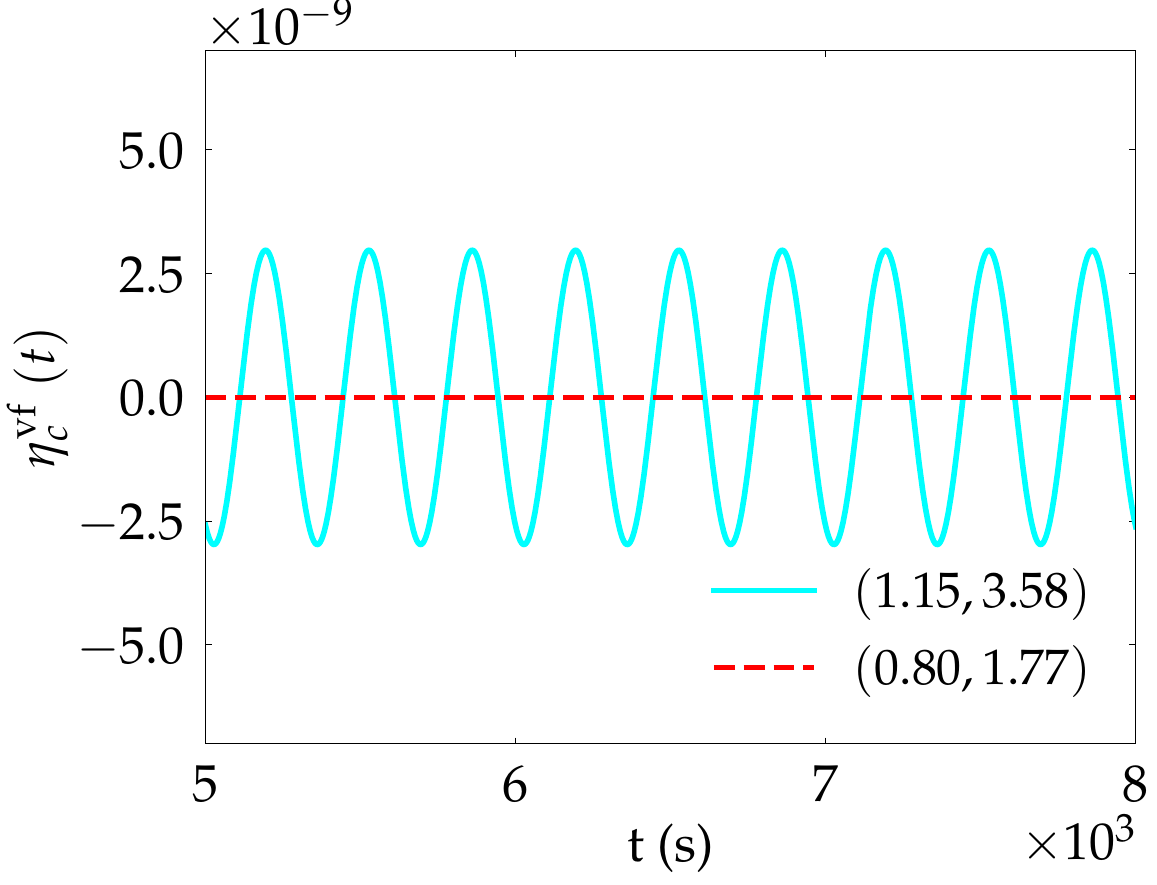}
    \caption{
    Time-domain signals in the vector NRC for $\hat{k}_c=\hat{k}$ (red-dashed) and $\hat{k}_c\neq\hat{k}$ (cyan-solid),
    generated by a transverse, circularly-polarized vector plane wave propagating in the direction $\hat{k}=(0.80,1.77)$ with $f=3~\text{mHz}$ and unit amplitude. The signal is strongly suppressed when the NRC is aligned with the source, i.e., $\hat{k}_c=\hat{k}$.}
    \label{fig:zs_vdm}
\end{figure}

\begin{figure}
    \centering
    \subfigure{
    \includegraphics[width=0.46\linewidth]{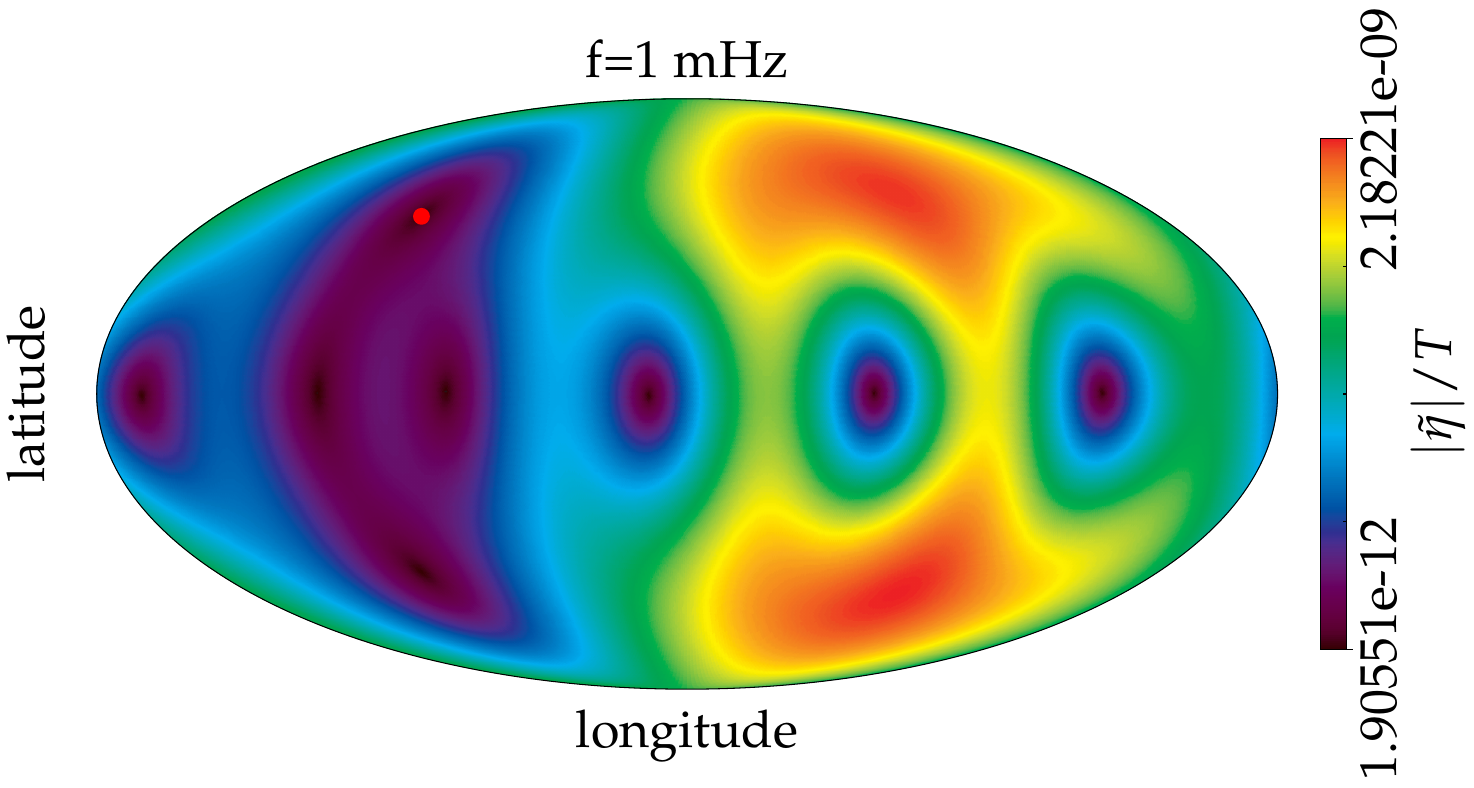}
    }
    \quad
    \subfigure{
    \includegraphics[width=0.46\linewidth]{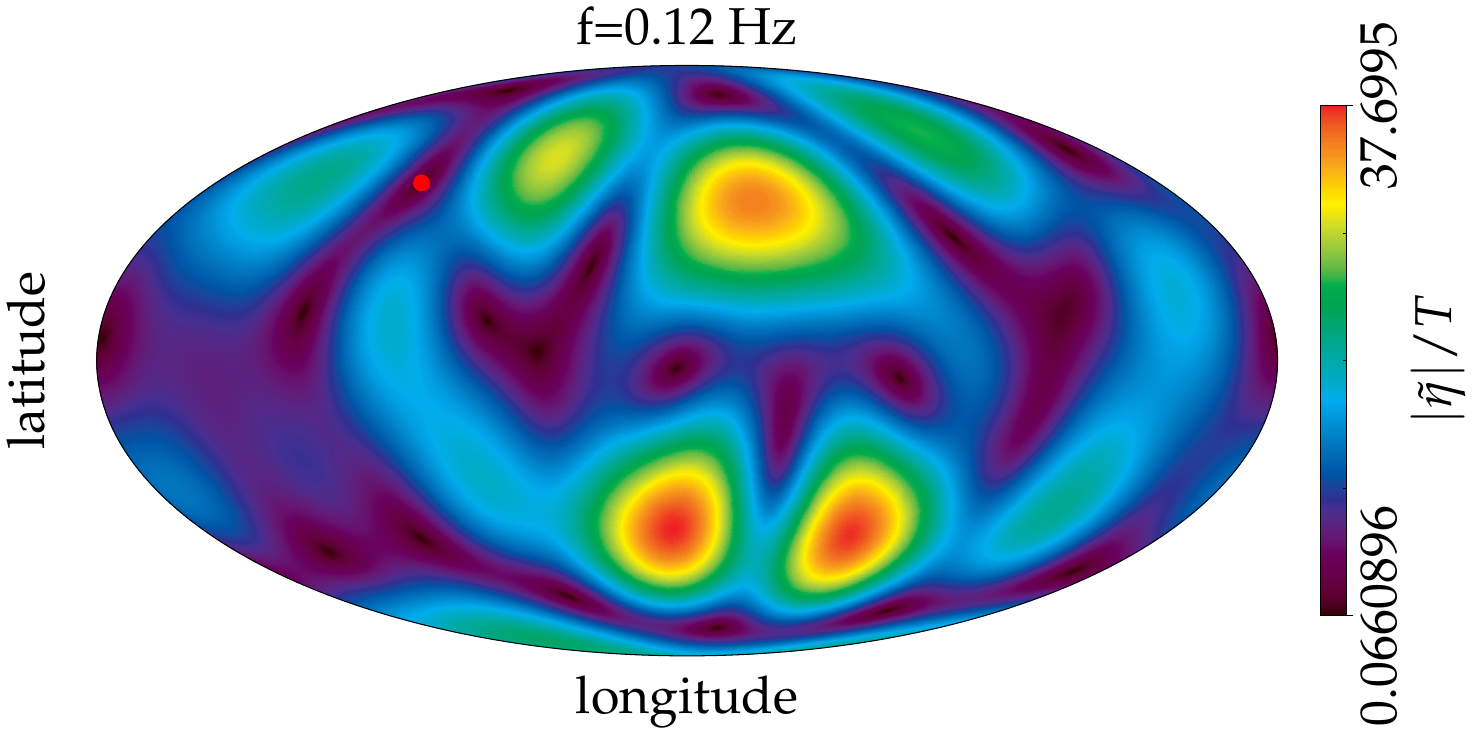}
    }
    \quad
    \subfigure{
    \includegraphics[width=0.46\linewidth]{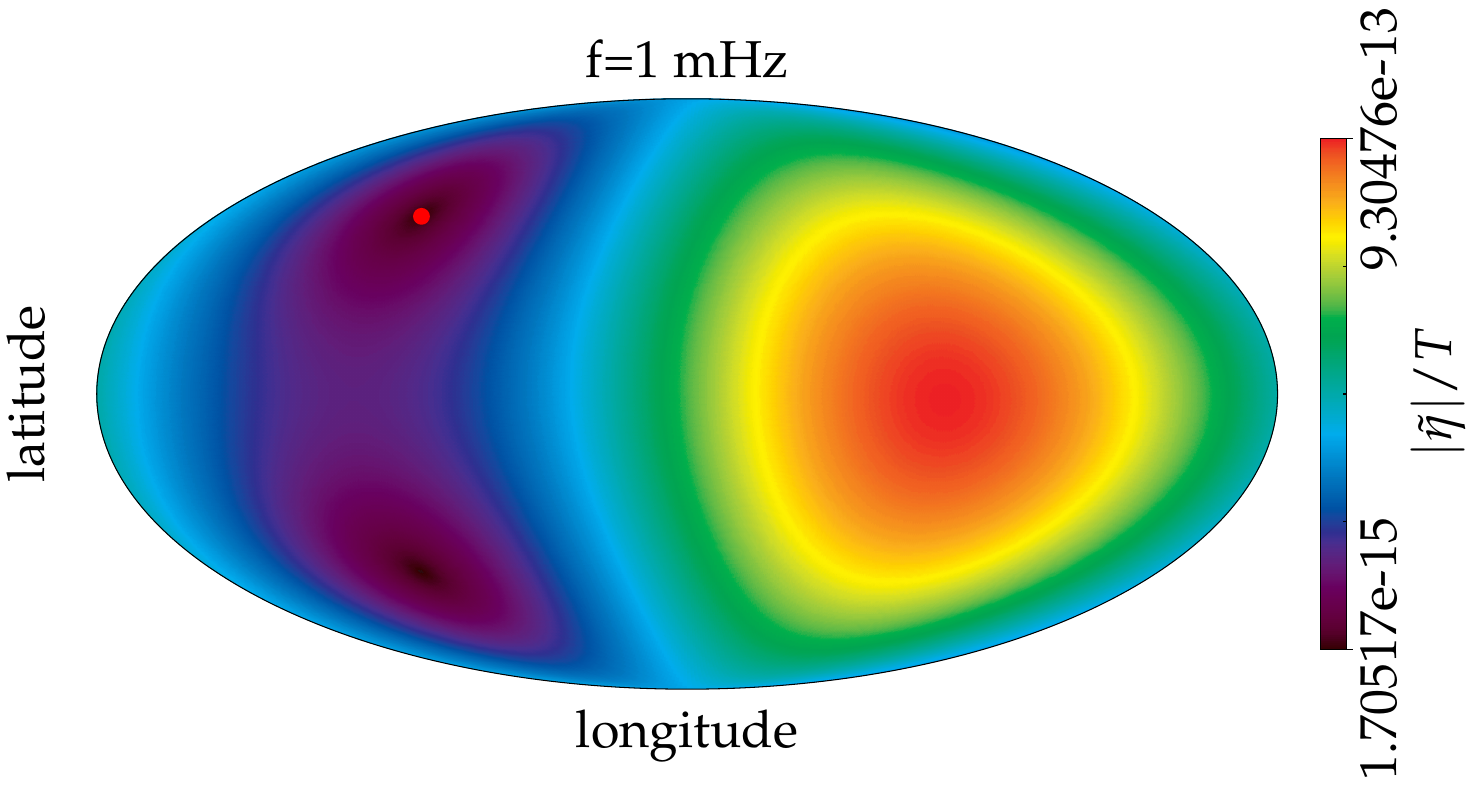}
    }
    \quad
    \subfigure{
    \includegraphics[width=0.46\linewidth]{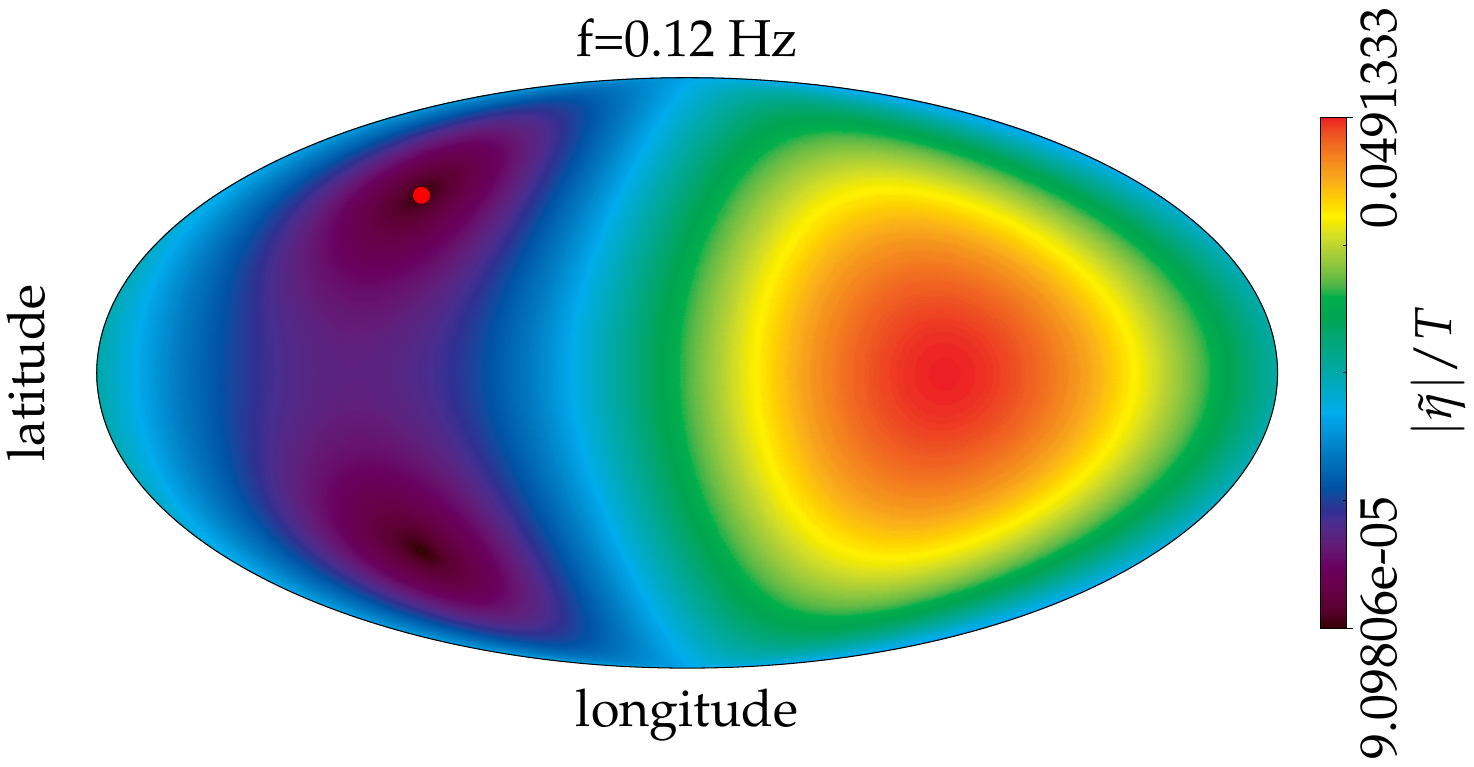}
    }
    \quad
    \subfigure{
    \includegraphics[width=0.46\linewidth]{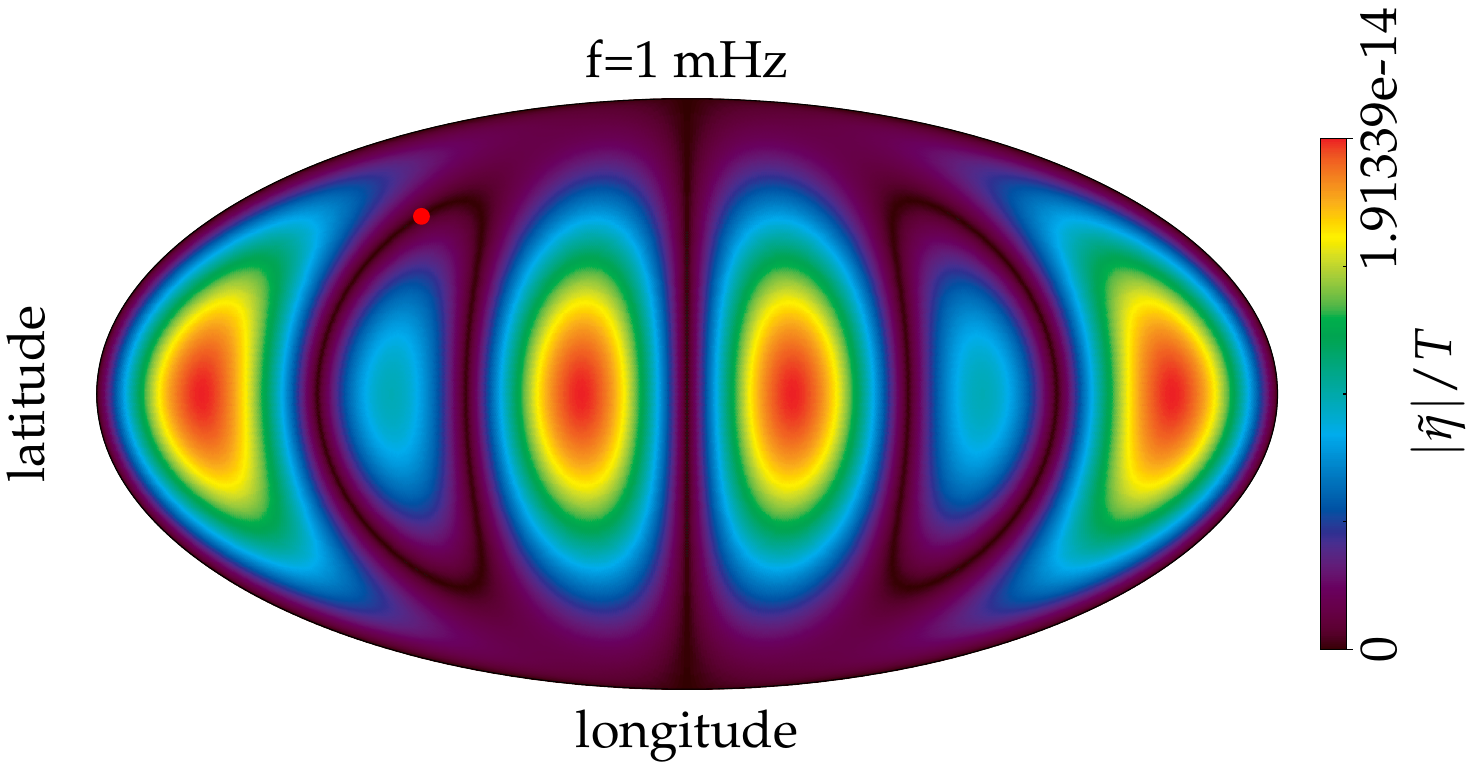}
    }
    \quad
    \subfigure{
    \includegraphics[width=0.46\linewidth]{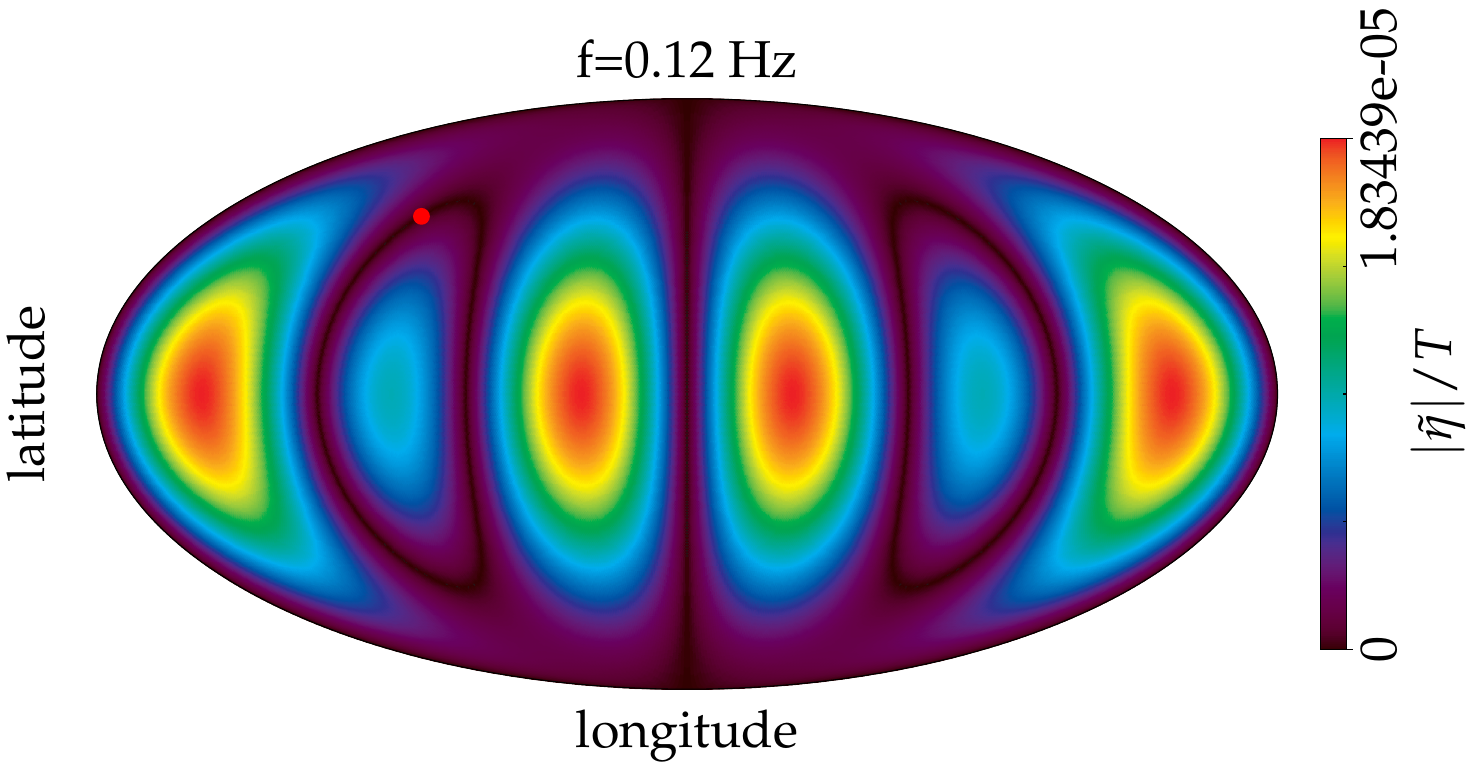}
    }
    \caption{Signal strength $|\tilde{\eta}|/T$ in NRCs for waves coming from various directions.
    From top to bottom: GW, vector, and scalar. Left for $f=1~\text{mHz}$ and right for $f=0.12~\text{Hz}$.
    The parameter $\hat{k}_c$ is fixed at $(0.80,1.77)$, marked by the red dot.
    Waves have unit amplitude.
    GWs and vector waves are assumed to be transverse and circularly-polarized. The package healpy~\cite{healpy, 2005ApJ...622..759G, Zonca2019} is used here to discretize the sky with $N_{\text{side}}=128$. 
    The point $(0,0)$ marks the center of each plot.}
    \label{fig:tsphere}
\end{figure}

In Fig.~\ref{fig:tsphere}, we show the strength of signals from various directions in their corresponding NRCs with a fixed parameter $\hat{k}_c$.
GWs and vector waves are assumed to be transverse and circularly-polarized.
Since the response is symmetric with respect to the detector plane,
signals propagating in the direction $(\theta,\phi) = (-\theta_c,\phi_c)$ are also strongly suppressed~\cite{Tinto:2004nz}.
Note that the strengths in the ULDM cases are symmetric across the entire band.
This is because, LISA-like detectors for the relevant mass range always operate in the long-wavelength limit ($2\pi L/\lambda \ll 1$), where the response to ULDM follows a simple dipole pattern.
In contrast, for GWs, the long-wavelength limit coincides with the low-frequency limit ($f \ll f_c \sim 1/2\pi L$), and the response reduces to a quadrupole pattern only within this regime.

\subsection{The NRC for a scalar ULDM field} \label{sec:NRC_sf}
Next we consider the scalar ULDM field that takes a plane-wave form
\begin{equation} \label{eq:scalar field}
    \Phi(x) = 
    \Phi_0e^{i(\omega t - \mathbf{k}\cdot \mathbf{x})}.
\end{equation}
As an illustration, here we consider the model where the scalar field is coupled through the trace of test mass's energy-momentum tensor~\cite{Morisaki:2018htj}.
Consequently, detectors are sensitive to the gradient of scalar field:
\begin{equation}
    \nabla\Phi(t,\mathbf{x}) = 
    -i\Phi_0e^{i(\omega t - \mathbf{k}\cdot \mathbf{x})} \mathbf{k},
\end{equation}
which behaves like a longitudinal-polarized vector field.

The scalar signal can be expressed as:
\begin{equation} \label{eq:eta sf}
    \tilde{\eta}_s = \left(\mathbf{a} \cdot \mathbf{x}^{\text{sf}}_l(\hat{k})\right) H_lT,
\end{equation}
where $H_l$ is the effective amplitude and $\mathbf{x}^{\text{sf}}_l(\hat{k}) = [\alpha^{\text{sf}}_l(\hat{k}), \beta^{\text{sf}}_l(\hat{k}), \gamma^{\text{sf}}_l(\hat{k})]^T$ with $\alpha^{\text{sf}}_l(\hat{k})$ the response of $\alpha$ combination to a longitudinal-polarized vector field.
Given that Eq.~(\ref{eq:eta sf}) involves only a single polarization mode, the scalar case admits multiple NRCs.
In fact, any coefficient vector in the plane orthogonal to $\mathbf{x}^{\text{sf}}_l(\hat{k})$ corresponds to a NRC.
Without loss of generality, we choose 
\begin{equation} \label{eq:NRC_sf_sp}
     \mathbf{a}^{\text{sf}}_c(\hat{k}_c) = [-\beta^{\text{sf}}_l(\hat{k}_c) - \gamma^{\text{sf}}_l(\hat{k}_c), \alpha^{\text{sf}}_l(\hat{k}_c), \alpha^{\text{sf}}_l(\hat{k}_c)]^T.
\end{equation}
The signal strength in the scalar NRC, Eq.~(\ref{eq:NRC_sf_sp}), for scalar waves arriving from various directions is shown in lowest row of  Fig.~\ref{fig:tsphere}.

Note that, since the gradient of a scalar field behaves like a longitudinal-polarized vector field, the vector NRC given by Eq.~(\ref{eq:NRC_vf}) also serves as a NRC for the scalar field model considered here. However, the reverse is not true---Eq.~(\ref{eq:NRC_sf_sp}) is not a NRC for a general vector field. This can be used to discriminate between different bosonic ULDM models.

\subsection{Angular resolution}
\begin{figure*}[t]
    \centering
    \includegraphics[width=0.48\linewidth]{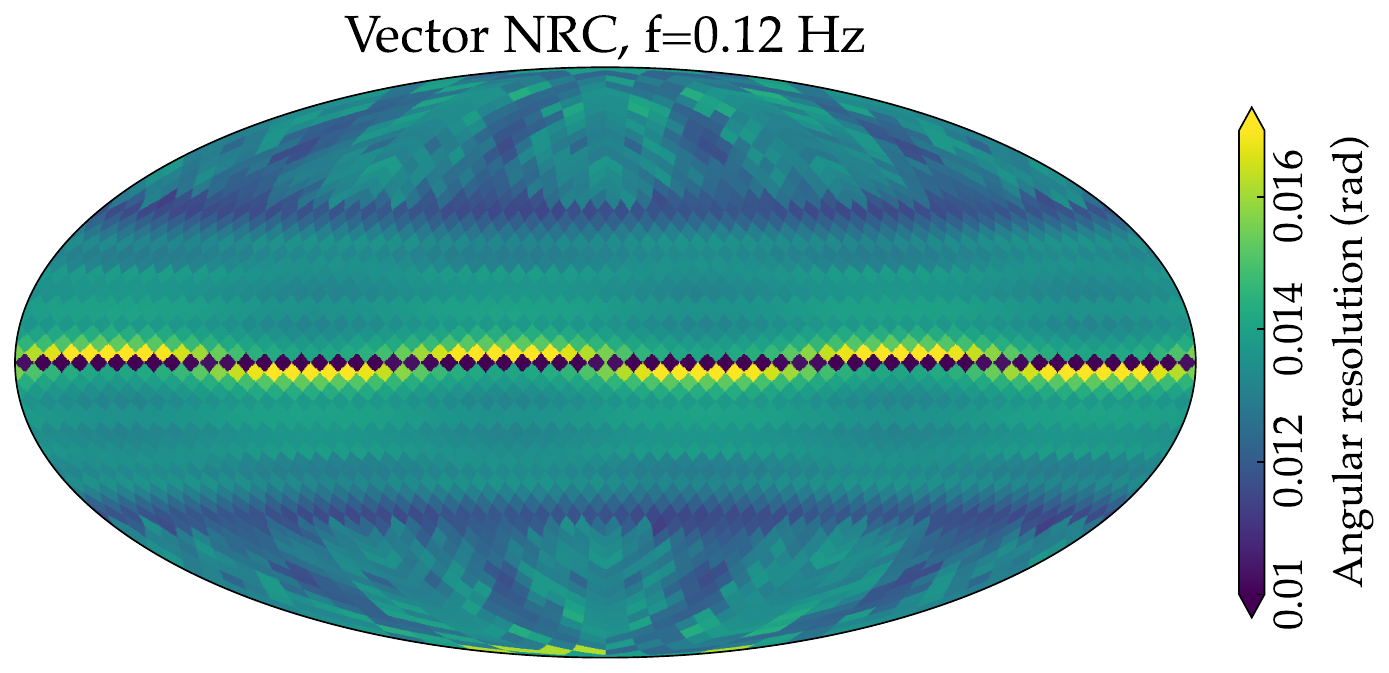}
    \includegraphics[width=0.48\linewidth]{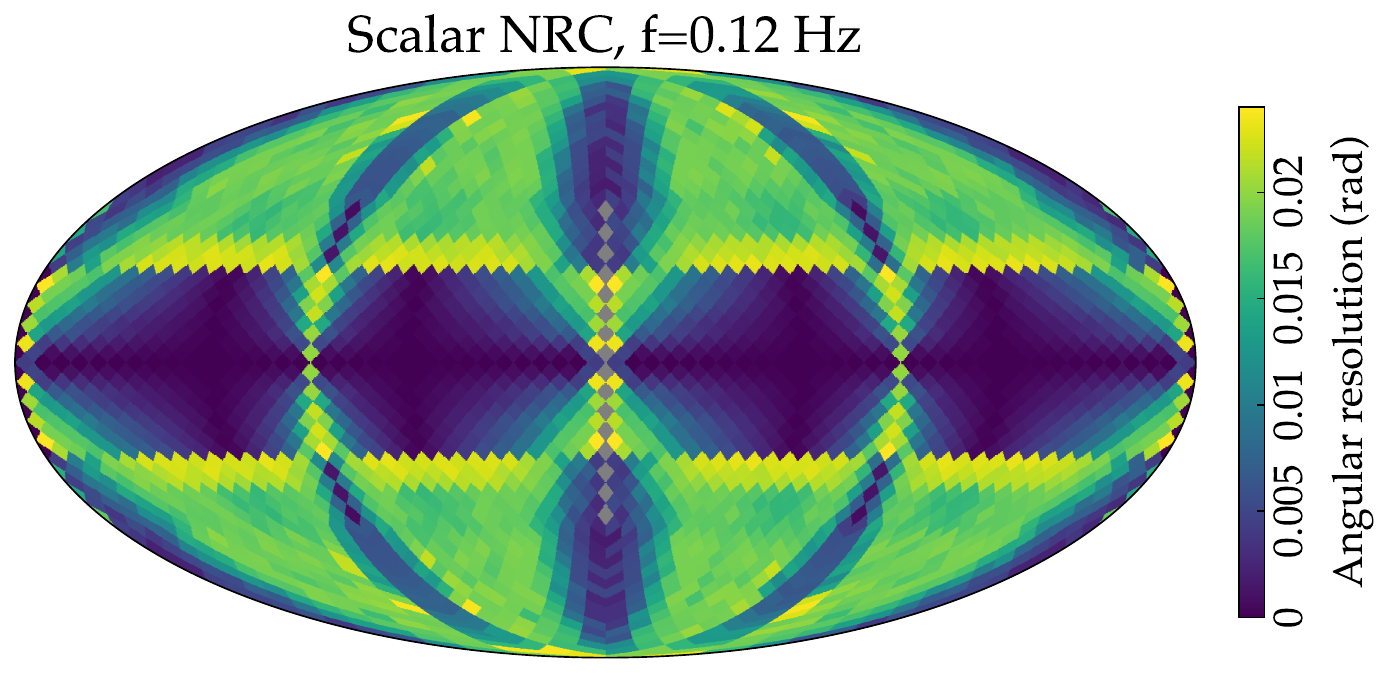}
    \caption{Angular resolution as a function of sky location. The vector field is assumed to be transverse and circular-polarized, the same as in Fig.~\ref{fig:tsphere}.}
    \label{fig:ar_angle}
\end{figure*}

We quantify the angular resolution of the NRCs using the gain function~\cite{Barroso:2024azl}, which is defined as the ratio of the median of the NRC response on a circle of radius $r$, centered at the reference point $(\theta_0, \phi_0)$ in $(\theta, \phi)$ plane, to that at the reference point, namely
\begin{equation}
    g(r) = 10 \log_{10}\frac{\text{median}|\tilde\eta^{\text{bf}}_c(\theta, \phi)|^2}{|\tilde\eta^{\text{bf}}_c(\theta_0, \phi_0)|^2} .
\end{equation}
The angular resolution is then defined as the radius with $g=3$, which means that the signal power from a direction offset by an angle $r$ is twice that from the reference point. We calculate the angular resolution over the full sky direction. The results are shown in Fig.~\ref{fig:ar_angle} for both vector and scalar cases. As shown, the angular resolution is $\mathcal{O}(10^{-2})$, at the same order as the case for GWs~\cite{Barroso:2024azl}.

\section{Discriminate ULDM and GW signals} \label{sec:apply}
\begin{figure}
    \centering
    \includegraphics[width=0.9\linewidth]{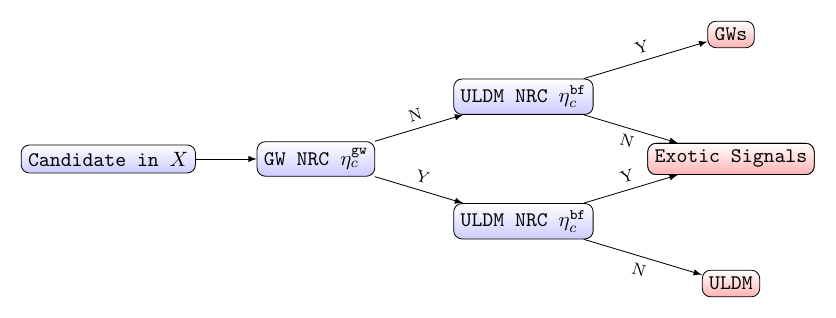}
    \caption{A schematic of the examination procedure. 
    ``Y" and ``N" represent a detectable signal and a null result, respectively.
    Note that if the outcome is ``YY" or ``NN", the exotic signal cannot be attributed to either GWs predicted by GR or the ULDM considered in this work.}
    \label{fig:exam}
\end{figure}

In this section, we explore the potential for distinguishing monochromatic signals produced by ULDM from those generated by GWs, using both the NRCs for GW and for ULDM.
The idea is as follows. Once a monochromatic signal is identified from the conventional Michelson-like $X$, we can construct the GW NRC and check the corresponding signal strength. Since GW NRC has zero response to GWs from a specific direction, a non-vanishing signal in this channel with $\hat{k}_c=\hat{k}$ would point to a non-GW origin---potentially ULDM. 
A similar reasoning applies to the NRC constructed for ULDM: if it yields a detectable signal, the source is unlikely to be ULDM. A schematic examination procedure is constructed here and summarized by Fig.~\ref{fig:exam}. Below we investigate the signal-to-noise (SNR) more quantitatively. 

\subsection{Employ the GW NRC to ULDM} \label{sec:gw2bf}
We first start with the ULDM signals observed in the GW NRC.
According to Eqs.~(\ref{eq:eta vf}), (\ref{eq:eta sf}) and (\ref{eq:NRC gw}), the signal is given by
\begin{equation}
    \tilde{\eta}^{\text{gw}}_{\text{bf}}(f,\hat{k};\hat{k}_c) = \sum_{p} \left(\mathbf{a}^{\text{gw}}_c(\hat{k}_c) \cdot \mathbf{x}^{\text{bf}}_{p}(\hat{k})\right) H_pT,
\end{equation}
where ``bf" stands for either scalar or vector field.
We define the response function, which translates the bosonic field amplitudes into the signal strength in GW NRC, as
\begin{equation} \label{eq:gw2dm}
    R^{\text{gw}}_{\text{bf},p}(f,\hat{k};\hat{k}_c) = 
    \mathbf{a}^{\text{gw}}_c(\hat{k}_c) \cdot \mathbf{x}^{\text{bf}}_{p}(\hat{k}) .
\end{equation}
We numerically evaluate the response function for various sets of parameters. 
We denote the magnitude of the scalar response function by
$R^{\text{gw}}_{\text{sf}}(f,\hat{k};\hat{k}_c) = \left|R^{\text{gw}}_{\text{sf},l}(f,\hat{k};\hat{k}_c)\right|$.
We define the polarization-averaged response function of vector field as
\begin{equation} \label{eq:polarization-averaged vf}
    R^{\text{gw}}_{\text{vf}}(f,\hat{k};\hat{k}_c) 
    = \sqrt{{\sum^3_{p=1} \left|R^{\text{gw}}_{\text{vf},p}(f,\hat{k};\hat{k}_c)\right|^2}\big{/}{3}}.
\end{equation}

\begin{figure*}[t]
    \centering
    \includegraphics[width=0.48\linewidth]{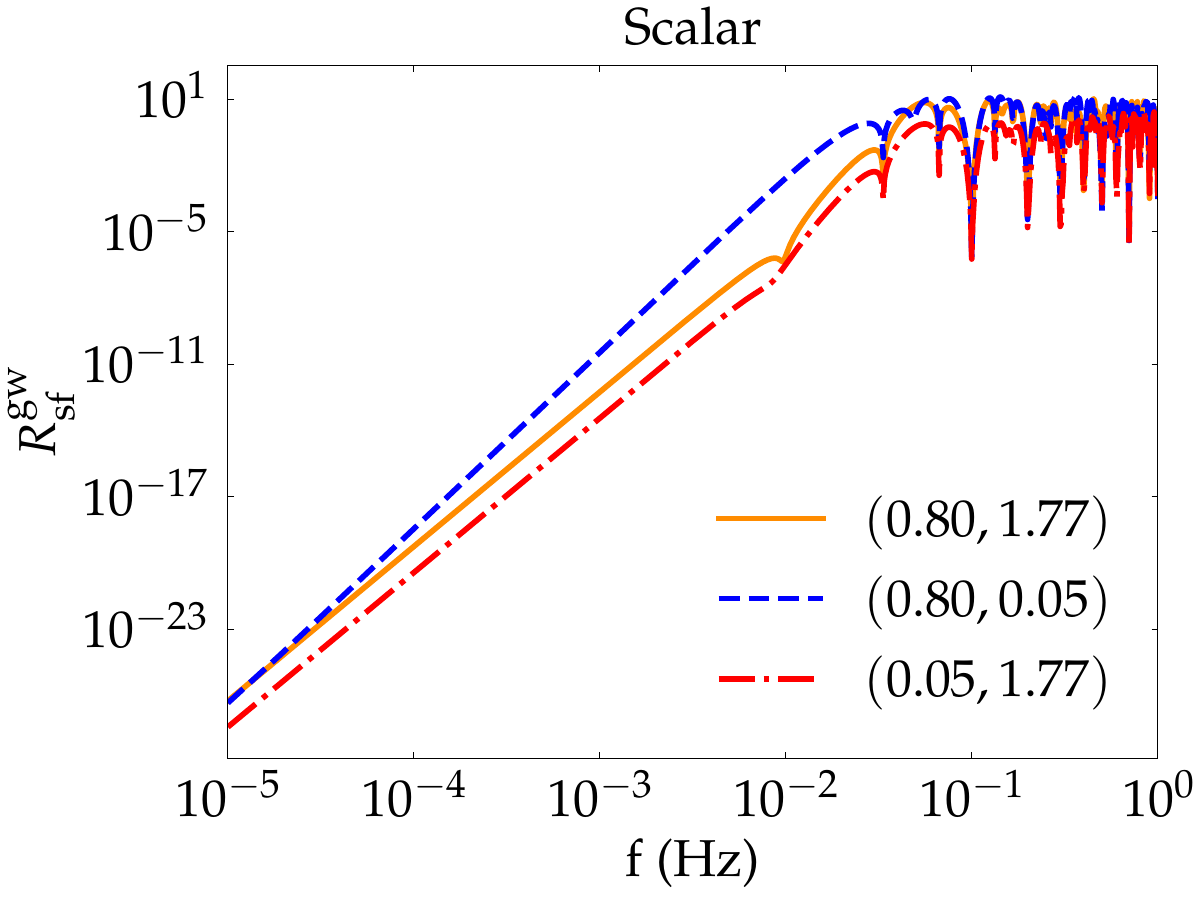}
    \includegraphics[width=0.48\linewidth]{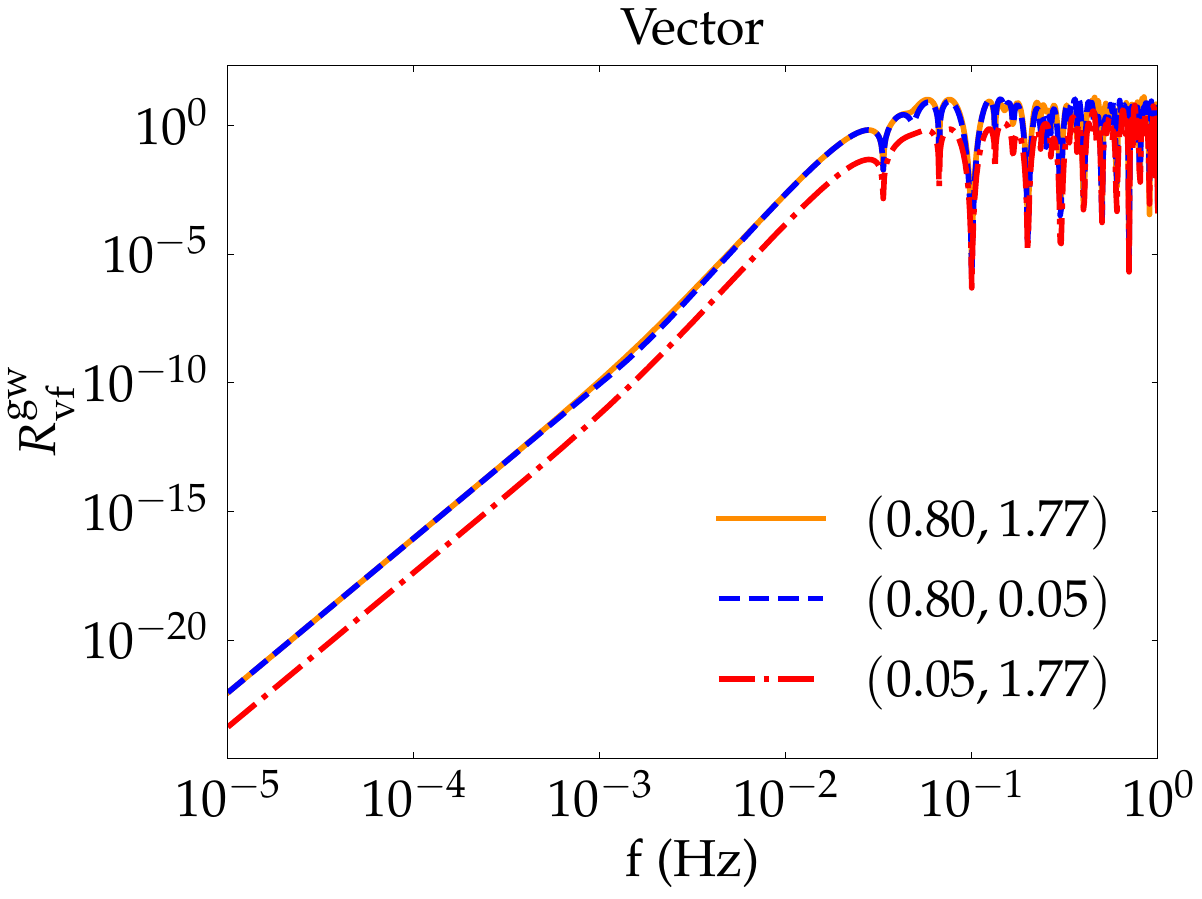}
    \caption{Response function for the ULDM signals in GW NRC. For vector, the polarization-averaged response function is plotted. The source location is fixed at $\hat{k}=(0.80,1.77)$, while we vary the parameter $\hat{k}_c$.}
    \label{fig:tf}
\end{figure*}

The response functions are shown in Fig.~\ref{fig:tf}. We fix $\hat{k} = (\theta,\phi) = (0.80, 1.77)$ while vary the parameter $\hat{k}_c = (\theta_c,\phi_c)$.
For the scalar field, the response function scales as $f^7$ in the low-frequency regime when $\phi_c = \phi$, while as $f^8$ when $\phi_c \neq \phi$.
The behavior of vector field differs from that of scalar field: 
$R^{\text{gw}}_{\text{vf}}$ scales as $f^6$ for $f<5\times10^{-4}~\text{Hz}$, and as $f^8$ at higher frequencies.

To validate our numerical result, we derive the response function in the low-frequency limit ($\delta = 2\pi f L \ll 1$). 
For scalar signal, it is given by
\begin{equation} \label{eq:gw2sf_lw}
\begin{split}
    R^{\text{gw}}_{\text{sf},l}(f,\hat{k};\hat{k}_c)
    \simeq
    & F(\hat{k}_c)\left[\frac{i}{2}\delta^7v
    [(\hat{k}_c\cdot \hat{n}_{1} )(\hat{k}\cdot\hat{n}_2 )(\hat{k}\cdot\hat{n}_3)
     + (1\rightarrow2\rightarrow3\rightarrow1)] \right.\\
    &\qquad\quad\left.-\frac{1}{4}\delta^8
    [(\hat{k}\cdot\hat{n}_1)(\hat{k}_c\cdot\hat{n}_2)-(\hat{k}\cdot\hat{n}_2)(\hat{k}_c\cdot\hat{n}_1)]
    + \mathcal{O}(\delta^{9}) \right],
\end{split}
\end{equation}
where $\hat{n}_i$ are the unit arm vectors defined in Fig.~\ref{fig:cor} and $F(\hat{k}_c)$ is given by
\begin{equation}
    F(\hat{k}_c)=\frac{3\sqrt{3}}{16}\left[\sin(3\theta_c)-7\sin\theta_c\right] .
\end{equation}
Note that, for $\hat{k}_c \neq \hat{k}$ and $\delta > v$, 
the leading contribution to $R^{\text{gw}}_{\text{sf},l}$ comes from the $\delta^8$ term. 
However, when $\hat{k}_c = \hat{k}$, the $\delta^8$ term vanishes, and the $\delta^7$ term becomes dominant. Additionally, since we choose the detector plane as the $x$-$y$ plane, the $\delta^8$ term also vanishes when $\phi_c = \phi$, as can be seen from Eq.~(\ref{eq:gw2sf_lw}).

The expression for $R^{\text{gw}}_{\text{vf},p}$ in the low-frequency limit is given by
\begin{equation} \label{eq:gw2vf_lw}
\begin{split}
    R^{\text{gw}}_{\text{vf},p} (f,\hat{k};\hat{k}_c)
    \simeq 
     -F(\hat{k}_c)
    &\left[\delta^6 v\left[(\hat{k}\cdot\hat{n}_1)(\hat{e}_p\cdot\hat{n}_2)-(\hat{k}\cdot\hat{n}_2)(\hat{e}_p\cdot\hat{n}_1)\right] 
     + \mathcal{O}(v\delta^7) \right.\\
    &\quad\left. +\frac{1}{4}\delta^8
    \left[(\hat{k}_c\cdot\hat{n}_1)(\hat{e}_p\cdot\hat{n}_2)-(\hat{k}_c\cdot\hat{n}_2)(\hat{e}_p\cdot\hat{n}_1)\right]
    + \mathcal{O}(\delta^9)
    \right].
\end{split}
\end{equation}
The frequency where $R^{\text{gw}}_{\text{vf}}$ changes its behavior from $f^8$ to $f^6$ is determined by $v\delta^6 \sim \delta^8$. For $L = 3\times10^6~\text{km}$, it is around $0.5~\text{mHz}$.
Thus, when $f < 5\times10^{-4}~\text{Hz}$, $R^{\text{gw}}_{\text{vf}}$ is dominated by the $v\delta^6$ term
and the response functions with different parameter $\hat{k}_c$ differ only by numerical factors given by $F(\hat{k}_c)$.
Note that, when set $\hat{e}_p = \hat{k}$, the $v\delta^6$ term vanishes and we recover Eq.~(\ref{eq:gw2sf_lw}).

While the GW NRC exhibits a non-zero response to ULDM, its SNR to a ULDM signal might be different from that in Michelson channel.
For monochromatic signal considered here, Eq.~(\ref{eq:eta SNR}) reduces to
\begin{equation}\label{eq:snr}
    \text{SNR}_{\eta} = \frac{S_{s}(f)}{S_n(f)},
\end{equation}
where the one-sided power spectral density~(PSD) of the ULDM signal observed in $\eta^{\text{gw}}_c$ is given by
\begin{equation} \label{eq:gw2bf signal PSD}
    S_s(f) = 2\frac{\left|\tilde{\eta}^{\text{gw}}_{\text{bf}}(f,\hat{k}_s;\hat{k})\right|^2}{T}
    = \left|\sum_p R^{\text{gw}}_{\text{bf},p}(f,\hat{k}_s;\hat{k}) H_p\right|^2 \frac{T}{2} ,
\end{equation}
where, in the final step, we account for the fact that the fields are represented by the real parts of Eq.~(\ref{eq:vec field}) and (\ref{eq:scalar field}).
For vector, we assume that the wave is linearly-polarized and average the signal over polarization directions and Eq.~(\ref{eq:gw2bf signal PSD}) reduces to 
\begin{equation} \label{eq:pav signal PSD}
    S_s(f) = \left(R^{\text{gw}}_{\text{vf}}(f,\hat{k}_s;\hat{k})\right)^2 S_h ,
\end{equation}
where we define the PSD of ULDM as $S_h = H^2/2T$ and $R^{\text{gw}}_{\text{vf}}$ is the polarization-averaged response function given by Eq.~(\ref{eq:polarization-averaged vf}).
For scalar, no averaging is needed and $H = H_l$.
The one-sided PSD of the noise is given by
\begin{equation}
    S_n(f) = \mathbf{a}^{\text{gw}}_c(f,\hat{k}_c)^{\dagger}\mathbf{S}_{\alpha}(f) \mathbf{a}^{\text{gw}}_c(f,\hat{k}_c) .
\end{equation}

\begin{figure*}[t]
    \centering
    \includegraphics[width=0.48\linewidth]{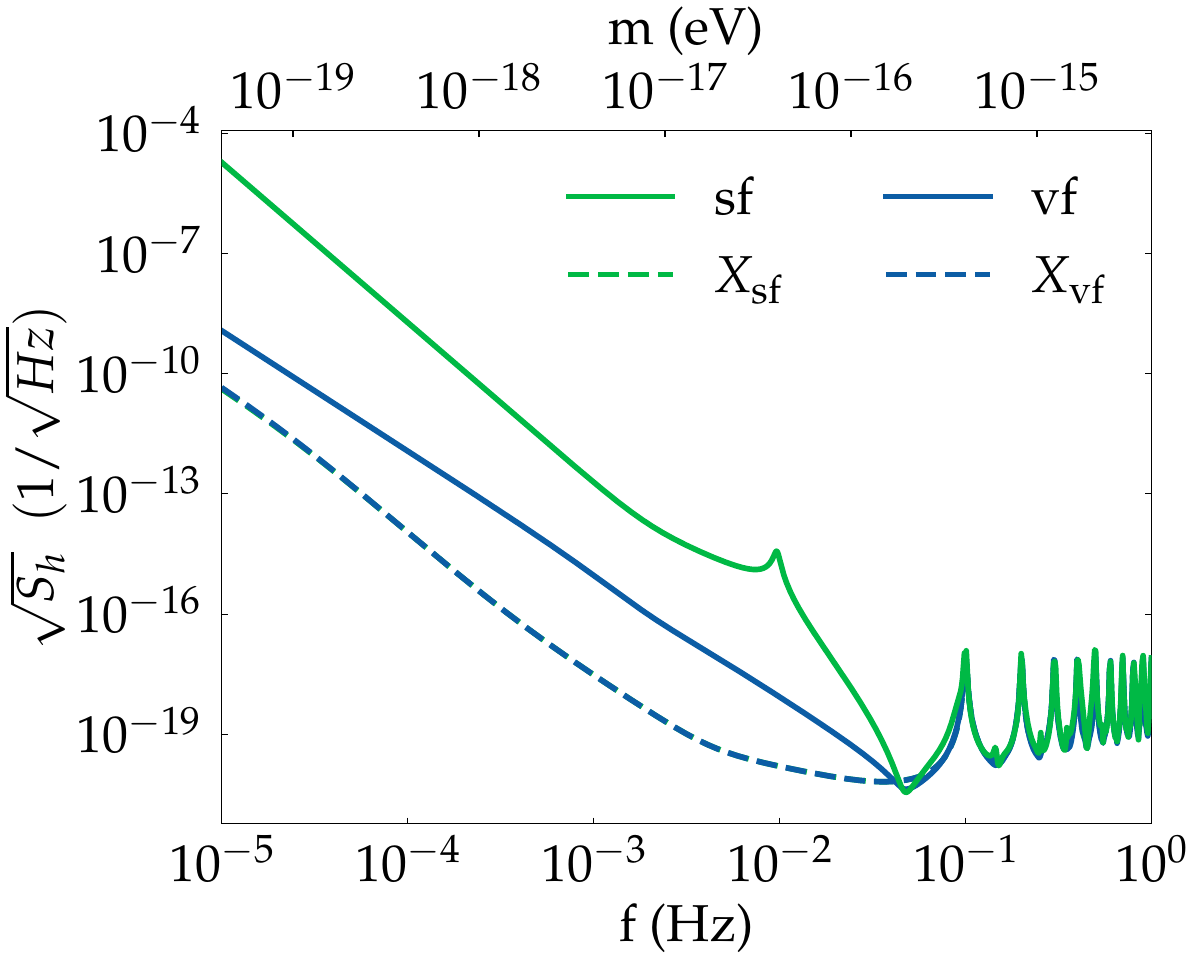}
    \includegraphics[width=0.48\linewidth]{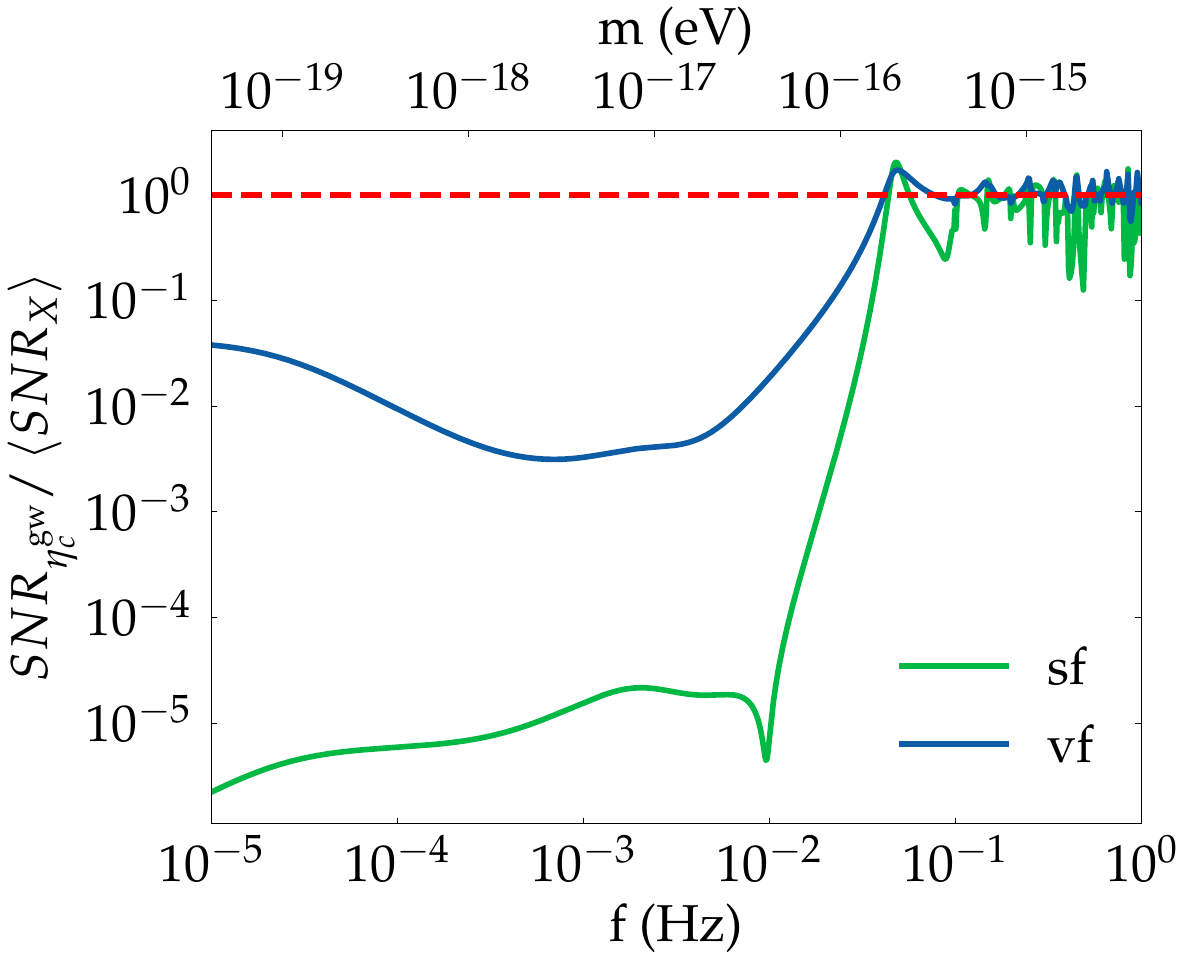}
    \caption{Left: The ULDM PSDs required to generate unit SNR in the GW NRC~(solid). Green for scalar and blue for vector. Also shown are the sky and polarization averaged PSDs in the $X$ combination (dashed), where the lines for scalar and vector are nearly overlapped.
    Right: The ratio between the SNR of the ULDM signal in the GW NRC and that in the $X$ combination. The red dashed line indicates where the ratio equal to one.}
    \label{fig:sen_gw2bf}
\end{figure*}

In Fig.~\ref{fig:sen_gw2bf}, we plot the ULDM PSD required to achieve a unit $\text{SNR}$ in the GW NRC, assuming $\hat{k}_c=\hat{k}=(0.8,1.77)$.
We also show the average PSD required to achieve a unit SNR for the Michelson-like $X$ combination~\cite{Yu:2023iog}.
As illustrated, while the GW NRC has a non-zero response to ULDM, its sensitivity is degraded in the low-frequency regime compared to the $X$ combination. 
For example, if a ULDM produces a monochromatic signal at $1~\text{mHz}$ with unit SNR in the $X$ combination,
the corresponding $\text{SNR}$ in $\eta^{\text{gw}}_c$ is $1.55\times10^{-5}$ and $3.27\times10^{-3}$ for scalar and vector, respectively.
However, in the high-frequency regime,
if a ULDM produces a signal at $50~\text{mHz}$ with unit SNR in the $X$ combination,
the corresponding $\text{SNR}$ in $\eta^{\text{gw}}_c$ increases to $1.9$ and $1.7$. Since $\eta^{\text{gw}}_c$ strongly suppresses GW signals from the same direction, the resulting $\text{SNR}$ for a GW would be vanishingly small, allowing one to distinguish a GW signal from a ULDM signal.

\subsection{Employ the ULDM NRC to GWs}
Now, we turn to the GW signals observed in the ULDM NRC.
We consider the signal generated by a monochromatic GW. The corresponding waveform is 
\begin{equation} \label{eq:gw mono waveform}
    \tilde{h}_p(f) = H_p\delta\left(f-f_{\text{gw}}\right) \simeq H_pT \delta_{f,f_{\text{gw}}}.
\end{equation}
The GW signal observed in ULDM NRC is given by
\begin{equation}
    \tilde{\eta}^{\text{bf}}_{\text{gw}}(f,\hat{k};\hat{k}_c) = \sum_{p=+,\times} \left(\mathbf{a}^{\text{bf}}_c(\hat{k}_c) \cdot \mathbf{x}^{\text{gw}}_{p}(\hat{k})\right) H_p T,
\end{equation}
where we have dropped the subscript in $f_{\text{gw}}$ without confusion.
The coefficient vector $\mathbf{a}^{\text{bf}}_c(\hat{k}_c)$ is given by
Eq.~(\ref{eq:NRC_sf_sp}) and (\ref{eq:NRC_vf}) for scalar and vector, respectively.
We define the response function as
\begin{equation}
    R^{\text{bf}}_{\text{gw},p}(f,\hat{k};\hat{k}_c) = 
    \mathbf{a}^{\text{bf}}_c(\hat{k}_c) \cdot \mathbf{x}^{\text{gw}}_{p}(\hat{k}).
\end{equation}
We numerically evaluate the response function for various sets of parameters. 
We fix $\hat{k} = (0.80, 1.77)$ as the location of GW source while vary the parameter $\hat{k}_c$. 
We define the polarization-averaged response function of GWs as:
\begin{equation} \label{eq:bf2gw pav}
    R^{\text{bf}}_{\text{gw}}(f,\hat{k};\hat{k}_c) 
    = \sqrt{{\sum_{p=+,\times} \left|R^{\text{bf}}_{\text{gw},p}(f,\hat{k};\hat{k}_c)\right|^2}\big{/}{2}}.
\end{equation}

\begin{figure}
    \centering
    \includegraphics[width=0.49\linewidth]{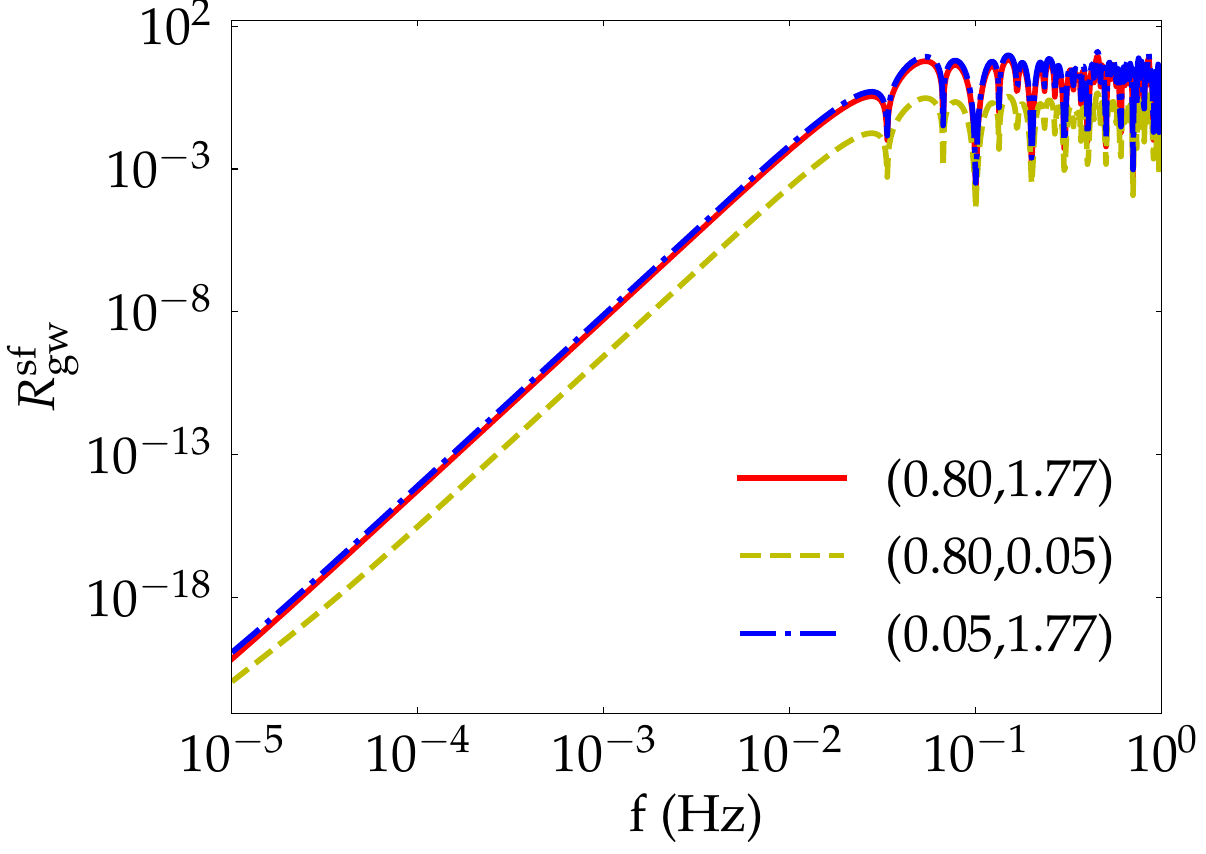}
    \includegraphics[width=0.49\linewidth]{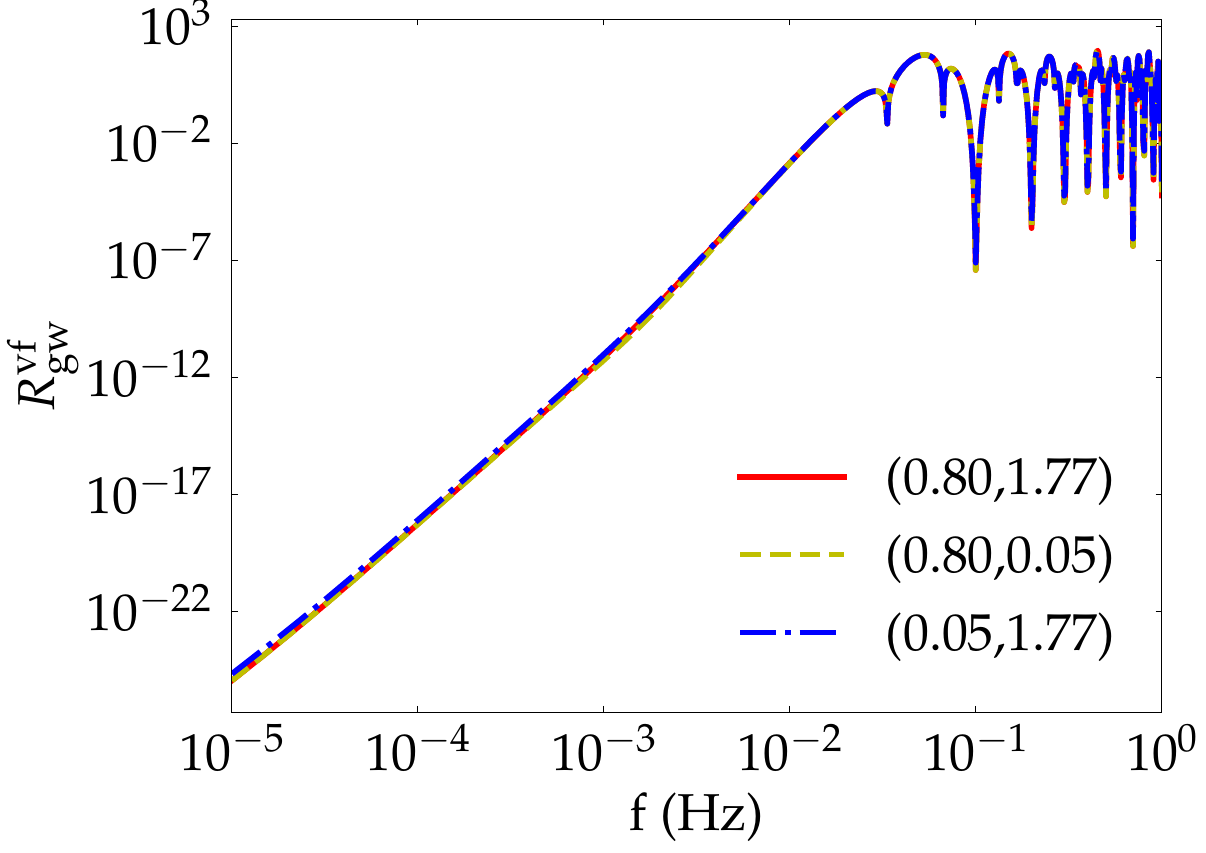}
    \caption{Response function for GW signals in the scalar (left) and vector (right) NRCs. The GW signals are averaged over polarizations, i.e., the polarization-averaged response function, Eq~(\ref{eq:bf2gw pav}), is plotted.
    The source location is fixed at $\hat{k}=(0.80,1.77)$, while we vary the parameter $\hat{k}_c$.}
    \label{fig:vf2gw}
\end{figure}

The response functions are shown in Fig.~\ref{fig:vf2gw}.
In the low-frequency regime, $R^{\text{vf}}_{\text{gw}}$ scales as $f^6$ for $f < 1.6\times 10^{-5}~\text{Hz}$, while as $f^7$ for $1.6\times 10^{-5}~\text{Hz}<f<10^{-3}~\text{Hz}$.
This behavior aligns with our expectation.
As reported in Ref.~\cite{Yu:2023iog}, the response of Sagnac combination to ULDM in the long-wavelength limit has an asymptotic behavior $vf^2 + f^3$ while that to GW is $f^2$.
Since the coefficient vector $\mathbf{a}^{\text{vf}}_c$ is quadratic in the response to ULDM,
the response function should behave as $(vf^2 + f^3)^2 \cdot f^2 \sim v^2f^6 + vf^7 + \mathcal{O}(f^8)$.
Explicit derivation shows that the $\mathcal{O}(f^8)$ terms cancel out each other and the leading velocity-independent term is $f^9$. 
Thus, in the low-frequency regime, we have $R^{\text{vf}}_{\text{gw}} \sim v^2f^6 + vf^7$.
Equating $v^2\delta^6$ with $v\delta^7$, we have $f \sim 1.6\times10^{-5}~\text{Hz}$, above which $R^{\text{vf}}_{\text{gw}}$ is dominated by the $vf^7$ term and below which the $v^2f^6$ term becomes important.

The response function $R^{\text{sf}}_{\text{gw}}$ scales as $f^5$ for $f < 1.6\times 10^{-5}~\text{Hz}$ and as $f^6$ for $1.6\times 10^{-5}~\text{Hz}<f<10^{-3}~\text{Hz}$.
Since the coefficient vector $\mathbf{a}^{\text{sf}}_c$, Eq.~(\ref{eq:NRC_sf_sp}), is linear in the response to ULDM, one should expect $R^{\text{sf}}_{\text{gw}} \sim (vf^2 + f^3) \cdot f^2 \sim vf^4 + f^5$. However, the sum of the coefficients in front of these terms yields zero. Consequently, 
$R^{\text{sf}}_{\text{gw}} \sim vf^5 + f^6$ in the low-frequency regime.

Notably, the response functions with different $\hat{k}_c$ seem only differ by an overall frequency-independent factor.
This can be understood as follow.
As mentioned earlier, for the nonrelativistic ULDM considered here, the long-wavelength approximation is valid.
Thus, the ULDM NRC reduces to the form $\mathbf{a}^{\text{bf}}_c(f,\hat{k}_c) \simeq \bar{\mathbf{a}}^{\text{bf}}_c(\hat{k}_c) \mathcal{T}(f)$, where $\mathcal{T}(f)$ only depends on frequency. This can be verified with the long-wavelength expressions of the response provided in the Appendix~\ref{ap:lfl_s}, but also can be seen directly from Fig.~\ref{fig:tsphere}, where the strength distribution remains unchanged for different frequencies in the ULDM case.

\begin{figure}
    \centering
    \includegraphics[width=0.48\linewidth]{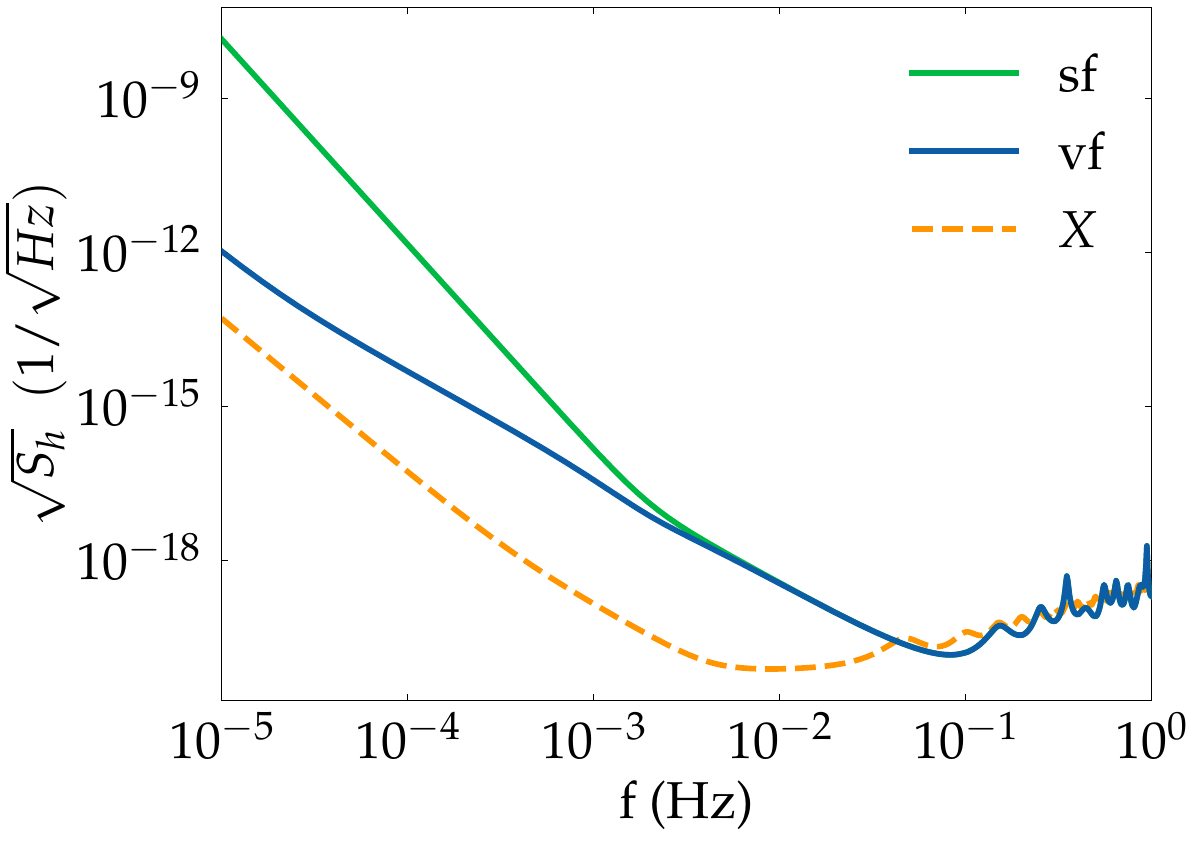}
    \includegraphics[width=0.48\linewidth]{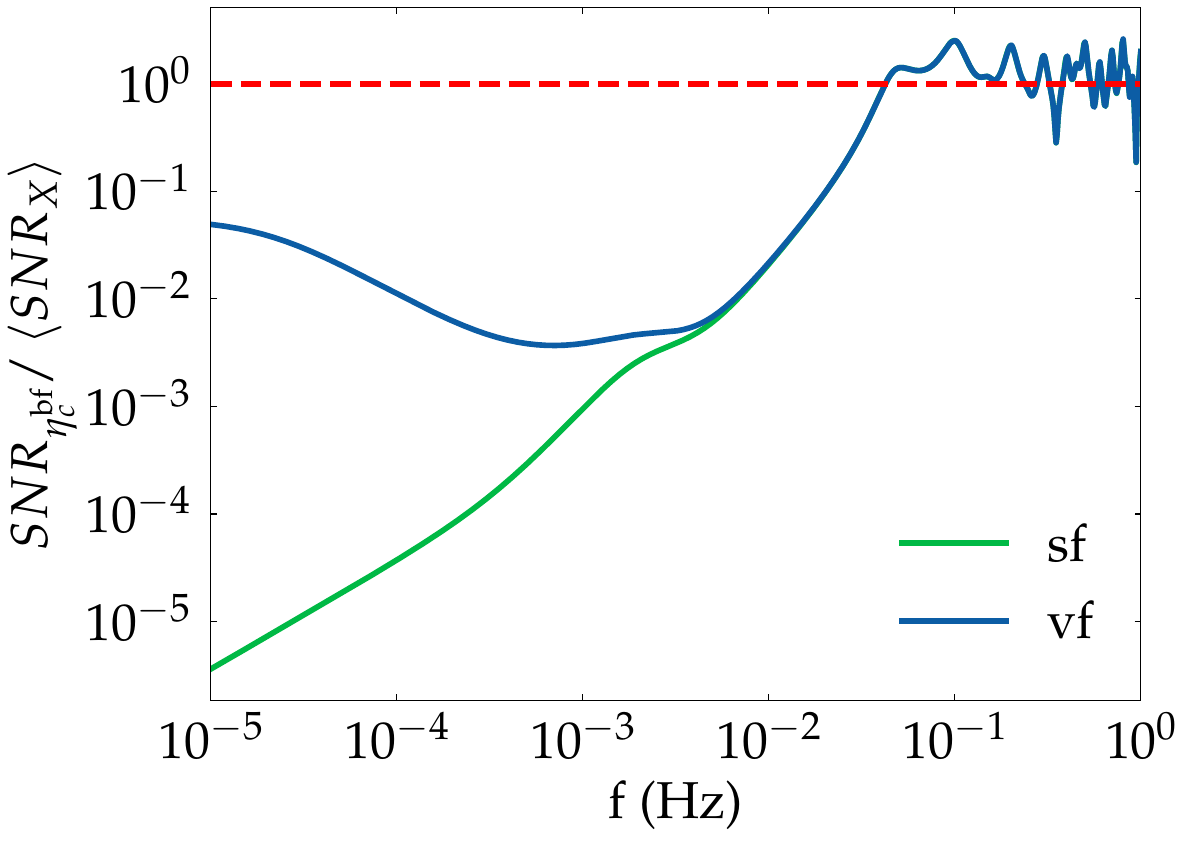}
    \caption{Left: The GW PSDs required to generate unit SNR in the ULDM NRCs~(solid). Also shown are the sky and polarization PSD in the $X$ combination.
    Right: The ratio between the SNR of GW signal in the ULDM NRCs and that in the $X$ combination. The dashed-red line indicates where the ratio equal to one.}
    \label{fig:sen_bf2gw}
\end{figure}

Aligning with the discussion in Sec.~\ref{sec:gw2bf}, we evaluate the PSD of GWs $S_h$ required to produce unit SNR in the ULDM NRCs, assuming $\hat{k}_c=\hat{k}=(0.8,1.77)$. Similar to Eq.~(\ref{eq:pav signal PSD}), we average the signal over GW polarizations. The results are shown in Fig.~\ref{fig:sen_bf2gw}, where the sky and polarization averaged PSD of the $X$ combination is also included.
As illustrated, while the ULDM NRCs is not sensitive to GW as Michelson channel at low frequencies, they can become more sensitive in the high-frequency regime.

\section{Discussion} \label{sec:limit}
In Sec.~\ref{sec:NRC_vf}, we consider a static detector and define the polarization modes with respect to the detector plane, which is crucial for constructing the vector NRC.
In practice, however, the three spacecrafts follow the heliocentric orbits and do not remain in a single plane over the course of the mission. 
Nevertheless, on a timescale of a few hundred seconds, the relative displacements between spacecraft are small enough that the constellation can be well approximated as lying in a common plane.
This allows us to divide the observation into short segments of such duration,
within which polarization modes can be defined relative to the instantaneous detector plane.
The corresponding NRC can then be constructed in each segment following the method developed in the main text. 
By gluing the segment-wise NRCs, we obtain a NRC which is valid over the entire mission duration.
Moreover, over a segment of this length, the relative motion between the sources and detector is negligible, thus the frequency modulation of signal can be ignored.

In this work, we model the ULDM field as a monochromatic plane wave.
However, since DM particles follow a velocity distribution, 
ULDM in reality should be described as a superposition of plane waves~\cite{PhysRevA.97.042506, PhysRevD.97.123006, Kim:2023pkx, Hui:2021tkt}.
Taking vector ULDM as an example, the field is described by
\begin{equation} \label{eq:DM random field}
    \mathbf{A}(x) = e^{imt}\sum_p a_p(x)  \hat{x}^p ,
\end{equation}
where $x=(t,\mathbf{x})$, $m$ is the mass of ULDM.
Temporal coherence and spatial coherence of the field are encoded in the complex amplitudes $a_p(x)$, which fluctuate stochastically over the coherence time and coherence length
\begin{equation}
    \tau_c = \frac{2\pi}{m\sigma^2} \approx 4.13 \times 10^8~\text{s}  \left(\frac{10^{-17}~\text{eV}}{m}\right),\quad
    \lambda_c = \sigma\tau_c \approx 1.24 \times 10^{11}~\text{km} \left(\frac{10^{-17}~\text{eV}}{m}\right),
\end{equation}
where $\sigma \sim 10^{-3}$ is the velocity dispersion of DM.
For scales shorter than $\tau_c$ and $\lambda_c$, the variation of $a_p(x)$ is negligible, and the amplitudes can be treated as constant. 
Since we only need to construct the NRC for a time segment of a few hundred seconds, which is much shorter than $\tau_c$, and the detector has a size of about $10^6~\text{km}$, which is much smaller than $\lambda_c$, the variation of the amplitudes can be safely ignored. Therefore, Eq.~(\ref{eq:DM random field}) coincides with a plane wave of $\mathbf{k}=0$. Thus, our formalism applies to the realistic ULDM field. 
 
Another important aspect of moving spacecraft is that, to suppress laser noise to the required level, one need employ the second-generation TDI~\cite{Tinto:2020fcc}, which has a  more complex algebraic structure than the first generation. 
Since the second generation has more than three generators, we expect our approach remains valid.
Moreover, the presence of multiple generators means that there are several inequivalent sets of generators that can be used to construct the NRC.
Which set yields a NRC most effective at distinguishing GW signals from those of ULDM remains an open question. 
We leave this problem for future work.

Finally, we note that in practice LISA tends to adopt a global-fit approach to simultaneously extract source and noise parameters~\cite{Littenberg:2023xpl}.
The problem we address here concerns identifying the physical origin of a monochromatic signal after its successful detection, and thus might be regarded as the second step after the global fit.
In this sense, the noise parameters used in our NRC analysis should be the output of the global fit.

\section{Conclusion} \label{Conclusions}
We have addressed the challenge of distinguishing monochromatic signals produced by gravitational waves~(GWs) from those potentially induced by ULDM in space-based GW detectors. Our approach leverages the null-response channel~(NRC), an interferometric combination designed to cancel signals from a specific source type at a given direction.  We generalize the NRC framework to accommodate multiple independent detectors and potential signals arsing from ULDM or additional polarizations in modified gravity. Using vector and scalar ULDM as examples, we construct their corresponding NRCs and analyze the responses to GWs. We also investigate how the GW NRC responds to ULDM.

Our results reveal that in the low-frequency regime the ULDM NRC strongly suppresses GW signals, while the GW NRC likewise suppresses ULDM signals. However, this changes in the high-frequency regime, where each NRC retains sensitivity to the other signal type at a level comparable to the conventional Michelson combination. This distinction enables a systematic procedure to discriminate between GW and ULDM signals, as illustrated in Fig.~\ref{fig:exam}. Our method provides a valuable tool for enhancing the scientific scope of future space-based GW detection missions.

\begin{acknowledgments}
This work is partly supported by the National Key Research and Development Program of China (Grant No.2021YFC2201901), 
the National Natural Science Foundation of China (Grant No.12147103),
and the Fundamental Research Funds for the Central Universities. H.T.Xu would like to thank R. Costa Barroso for helpful communications. 
\end{acknowledgments}

\textbf{Data availability}: The data that support the findings of this article are openly available~\cite{xu_2025_17448354}.

\appendix
\section{Signal Response} \label{ap:lfl_s}
The response to a field's polarization mode $p$ is defined as 
the Fourier transform of the signal, which is generated by a monochromatic $p$-polarized plane wave with $H_p=1$ and all other components set to zero, divided by the Dirac delta function arising from the Fourier transform of a monochromatic signal.
For example, the single-link signal generated by a $p$-polarized monochromatic GW with unit amplitude is given by
\begin{equation} \label{eq:1link gw mono}
    y^{\text{gw},p}_{ij}(t) = 
    \frac{\hat{n}_{ij}\otimes \hat{n}_{ij}:\mathbf{e}^{p}}{2(1-\hat{k}\cdot \hat{n}_{ij})}\left[e^{-i\omega(L_{ij}+\hat{k}\cdot\mathbf{x}_j)}-e^{-i\omega\hat{k}\cdot \mathbf{x}_i}\right]e^{i\omega t},
\end{equation}
where
\begin{equation}
    \mathbf{e}^{+} = \hat{u}\otimes\hat{u} - \hat{\nu}\otimes\hat{\nu}, \qquad
    \mathbf{e}^{\times} = \hat{u}\otimes\hat{v} + \hat{v}\otimes\hat{u},
\end{equation}
where $\hat{u} = (\cos\theta)^{-1}\partial\hat{k}/\partial\phi$ and $\hat{v} = -\partial\hat{k}/\partial\theta$.
The Fourier transform of Eq.~(\ref{eq:1link gw mono}) is given by
\begin{equation} 
    \tilde{y}^{\text{gw},p}_{ij}(f) = y^{\text{gw},p}_{ij} \delta(f-f_{\text{gw}}),
\end{equation}
where the single-link response is given by
\begin{equation}
    y^{\text{gw},p}_{ij}(f,\hat{k}) = 
    \frac{\hat{n}_{ij}\otimes \hat{n}_{ij}:\mathbf{e}^{p}}{2(1-\hat{k}\cdot \hat{n}_{ij})}\left[e^{-i2\pi f(L_{ij}+\hat{k}\cdot\mathbf{x}_j)}-e^{-i2\pi f\hat{k}\cdot \mathbf{x}_i}\right] .
\end{equation}
Similarly, the single-link response to ULDM can be obtained by Fourier transforming Eq.~(\ref{eq:1link bf}).

The expression for Sagnac $\alpha$ combintion is given in Eq.~(\ref{eq:def sagnac}) and
those for $\beta$,$\gamma$ can be obtained by cycling the index.
The symmetric Sagnac combination is defined as
\begin{equation}
    \zeta(t) = y_{21,2}-y_{12,1}+y_{13,1}-y_{31,3}+y_{32,3}-y_{23,2}.
\end{equation}
In the Fourier domain, the delay operation corresponds to multiplication by the appropriate delay factor. For example, 
\begin{equation}
    y_{ij,k}(t) = y_{ij}(t - L_k) \longleftrightarrow
    \tilde{y}_{ij,k}(f) = e^{-i2\pi fL_k}\tilde{y}_{ij}(f).
\end{equation}
Therefore, the responses of Sagnac combinations can be obtained by 
linearly combining the single-link responses with the corresponding delay factors as coefficients.

While the general expressions for the responses are complicated, they can be expanded in powers of $\delta = 2\pi fL$ in the low-frequency regime.
For GW, the Sagnac responses can be approximated as
\begin{align}
    \alpha^{\text{gw}}_p(f,\hat{k}) & \simeq \delta^2\left[\hat{n}_2 \cdot \mathbf{e}^p(\hat{k}) \cdot \hat{n}_2-\hat{n}_3 \cdot \mathbf{e}^p(\hat{k})\cdot \hat{n}_3\right],\\
    \zeta^{\text{gw}}_p(f,\hat{k}) &\simeq -\frac{i\delta^3}{12}\left[(\hat{k}\cdot\hat{n}_1)(\hat{n}_1 \cdot \mathbf{e}^p(\hat{k}) \cdot \hat{n}_1) + (1\rightarrow2\rightarrow3)\right],\label{eq:gwzeta}    
\end{align}
where $\hat{n}_i$ are the unit arm direction vectors defined in Fig.~\ref{fig:cor}.
For bosonic fields, the low-frequency approximation are given by
\begin{align}
    \alpha^{\text{bf}}_p(f,\hat{k}) & \simeq v\delta^2[(\hat{e}_p\cdot \hat{n}_2)(\hat{k}\cdot\hat{n}_1) - (\hat{e}_p\cdot\hat{n}_1)(2\hat{k}\cdot\hat{n}_3-\hat{k}\cdot \hat{n}_2)] + i\delta^3\hat{e}_p\cdot\hat{n}_1,\label{eq:bfalpha}\\
    \zeta^{\text{bf}}_p(f,\hat{k}) &\simeq \frac{v\delta^2}{3}\left[\left((\hat{e}_p\cdot \hat{n}_1)(\hat{k}\cdot\hat{n}_2)-(\hat{e}_p\cdot\hat{n}_2)(\hat{k}\cdot\hat{n}_1)\right) + (1\rightarrow2\rightarrow3)\right]
    ,\label{eq:bfzeta}
\end{align}
where we use the geometric identity $\sum_i\hat{n}_{i} = 0$ in the derivation.
For scalar fields where $\hat{e}_p = \hat{k}$, we see that the $v\delta^2$ term in Eq.~(\ref{eq:bfzeta}) vanishes and the leading term is 
\begin{equation}\label{eq:sfzeta}
    \zeta^{\text{sf}}_l(f) \simeq -\frac{i}{2}v^2\delta^3(\hat{k}\cdot \hat{n}_1)(\hat{k}\cdot \hat{n}_2)(\hat{k}\cdot \hat{n}_3) .
\end{equation}

Using the low-frequency expressions of Sagnac combinations, we derive the responses of NRC to the corresponding signal.
For vector, 
\begin{align}\label{eq:NRC_bf}
     \eta^{\text{vf}}_{c,p}(f,\hat{k};\hat{k}_c)\simeq & 3v\delta^8 
     \left(F^x_2 F^y_3-F^y_2 F^x_3\right) 
      \left(\hat{k} - \hat{k}_c\right)
      \left[ \left(\hat{n}_{2}\cdot\hat{e}_p\right)\hat{n}_{1} - \left(\hat{n}_{1}\cdot\hat{e}_p\right)\hat{n}_{2}\right],
\end{align}
where $F^{x}_i = \hat{x}\cdot\hat{n}_i$ and $F^{y}_i = \hat{y}\cdot\hat{n}_i$.
For scalar,
\begin{align}\label{eq:NRC_sf}
    \eta^{\text{sf}}_{c,l}(f,\hat{k};\hat{k}_c)
    \simeq \frac{3}{2} v^2\delta^6(\hat{k}_c\cdot\hat{n}_1)(\hat{k}\cdot\hat{n}_1)
    \left[(\hat{k}\cdot \hat{n}_2)(\hat{k}\cdot\hat{n}_3)-(\hat{k}_c\cdot \hat{n}_2)(\hat{k}_c\cdot\hat{n}_3)\right],
\end{align}
For GWs,
\begin{align}\label{eq:NRC_gw}
    \eta^{\text{gw}}_{c,p}(f,\hat{k};\hat{k}_c) \simeq & -\frac{i}{4}\delta^7 F(\hat{k}_c)(\hat{k}-\hat{k}_c)\cdot\left(\sum_i\hat{n}_iF^{p}_i(\hat{k}) \right),
\end{align}
where $F^{p}_i(\hat{k}) = \hat{n}_i \cdot\mathbf{e}^p(\hat{k}) \cdot \hat{n}_i$ and
\begin{equation}
    F(\hat{k}) = 4[(F^{+}_1 F^{\times}_2-F^{\times}_1 F^{+}_2)+(F^{+}_2 F^{\times}_3-F^{\times}_2 F^{+}_3)+(F^{+}_3 F^{\times}_1-F^{\times}_3 F^{+}_1)].
\end{equation} 
Using the above analytic expressions, we can recover the sky pattern shown in Fig.~\ref{fig:tsphere}.

\section{The NRC Coefficients of Vector Field} \label{ap:coefficients}
We define the polarization modes relative to the coordinate system in Fig.~\ref{fig:cor}.
The responses to the two transverse polarizations ($\hat{e}_p = \hat{x}, \hat{y}$) are given by
\begin{equation}
    \alpha^{\text{vf}}_{x,y}(f,\hat{k})=\alpha_1 F^{x,y}_1+\alpha_2 F^{x,y}_2+\alpha_3 F^{x,y}_3,
\end{equation}
where $F^{p}_i = \hat{e}_p \cdot \hat{n}_i$ and
\begin{align}
    \alpha_1&=[e^{-i2\pi f (L_2+\hat{k}\cdot \vec{x}_3v)}-e^{-i2\pi f(L_2+L_1+\hat{k}\cdot \vec{x}_2v)}]+[e^{-i2\pi f(L_3+\hat{k}\cdot \vec{x}_2 v)}-e^{-i2\pi f(L_3+L_1+\hat{k}\cdot \vec{x}_3 v)}],\\
    \alpha_2&=[e^{-i2\pi f(\hat{k}\cdot \vec{x}_1v)}-e^{-i2\pi f(L_2+\hat{k}\cdot \vec{x}_3 v)}]+[e^{-i2\pi f(L_1+L_3 +\hat{k}\cdot \vec{x}_3 v)}+e^{-i2\pi f(L_1+L_2+L_3+\hat{k}\cdot \vec{x}_1 v)}],\\
    \alpha_3 &=[e^{-i2\pi f(L_1+L_2+\hat{k}\cdot \vec{x}_2v)}-e^{-i2\pi f(L_1+L_2+L_3+\hat{k}\cdot \vec{x}_1 v)}]+[e^{-i2\pi f(\hat{k}\cdot \vec{x}_1 v)}-e^{-i2\pi f(L_3+\hat{k}\cdot \vec{x}_2 v)}].
\end{align}
The expressions for $\beta$ and $\gamma$ can be obtained by permuting the indices.

The coefficient vector $\mathbf{a}^{\text{vf}}_c$ can be derived straightforwardly. 
For example, 
\begin{equation}
    a_3 = [F_2^x F_3^y - F_2^y F_3^x] \sum_{k=1}^{25}B_3^{(k)}e^{-2\pi i f \Delta_3^{(k)}},
\end{equation}
where we used $F_1^x F_2^y-F_1^y F_2^x=F_3^x F_1^y - F_3^y F_1^x = F_2^x F_3^y - F_2^y F_3^x$ which follows from the identity $\sum_i\hat{n}_{i}= 0$.
The $25$ time delays $\Delta_3^{(k)}$ are collected in Table~\ref{delta}, and the coefficients $B_3^{(k)}$ are given by
\begin{equation}
\begin{split}
     &B_3^{(1)}=B_3^{(2)}=B_3^{(5)}=B_3^{(6)}=2,\\
     &B_3^{(3)}=B_3^{(4)}=B_3^{(7)}=B_3^{(8)}=-2,\\
     &B_3^{(14)}=B_3^{(17)}=B_3^{(20)}=B_3^{(23)}=-1,\\
     &B_3^{(15)}=B_3^{(16)}=B_3^{(21)}=B_3^{(22)}=-3,\\
     &B_3^{(18)}=B_3^{(19)}=B_3^{(24)}=B_3^{(25)}=4,\\
     &B_3^{(9)}=B_3^{(12)}=1,\\
     &B_3^{(10)}=B_3^{(11)}=3,\\     
     &B_3^{(13)}=-8.
\end{split}
\end{equation}

Assuming equal arm lengths $L_i = L_j = L$ and set $v=0$, we have
\begin{equation}
    a_1 = a_2 = a_3 = e^{6i\delta}\left(1-e^{i\delta}\right)^6[F_2^x F_3^y - F_2^y F_3^x].
\end{equation}
Consequently, $\tilde{\eta}^{\text{vf}} \propto (\tilde{\alpha} + \tilde{\beta} + \tilde{\gamma})$.
There is a relationship between $\zeta$ and other Sagnac combinations:
\begin{equation}\label{eq:zeta}
    \left(1 - e^{-3i\delta}\right)\tilde{\zeta} = e^{-i\delta}\left(1 - e^{-i\delta}\right)(\tilde{\alpha} + \tilde{\beta} + \tilde{\gamma}).
\end{equation}
Thus, we have $\tilde{\eta}^{\text{vf}} \propto \tilde{\zeta}$ for $e^{-i\delta} \neq 1$.
According to Eq.~(\ref{eq:eta SNR}), combinations that differ only by an overall factor have the same SNR.

\begin{table}
\caption{The 25 time delays entering into the function $a_3$. The other 50 time delays, entering into the functions $a_1$ and $a_2$, can be obtained by permuting the indices.}
\label{delta}
\resizebox{\textwidth}{!}{
\begin{tabular}{ccccccccccc}
\hline
$k$ & & $\Delta_3^{(k)}$ & & $k$ & & $\Delta_3^{(k)}$ & & $k$ & & $\Delta_3^{(k)}$  \\
\hline
1 & & $L_1+L_2+2\hat{k}\cdot \vec{x}_1v$ & & 10 & & $2L_1+2L_2+\hat{k}\cdot \vec{x}_1v+\hat{k}\cdot \vec{x}_2v$ & & 19  & & $2L_1+L_2+L_3+\hat{k}\cdot \vec{x}_1v+\hat{k}\cdot \vec{x}_3v$ \\
2 & & $L_1+L_2+2L_3+2\hat{k}\cdot \vec{x}_1v$ & & 11 & & $2L_3+\hat{k}\cdot \vec{x}_1v+\hat{k}\cdot \vec{x}_2v$ & & 20 & & $L_2+\hat{k}\cdot \vec{x}_2v + \hat{k}\cdot \vec{x}_3v$ \\
3 & & $L_3+2\hat{k}\cdot \vec{x}_1v$ & & 12 & & $2L_1+2L_2+2L_3+\hat{k}\cdot \vec{x}_1v+\hat{k}\cdot \vec{x}_2v$ & & 21 & & $2L_1+L_2+\hat{k}\cdot \vec{x}_2v+\hat{k}\cdot \vec{x}_3v$ \\
4 & & $2L_1+2L_2+L_3+2\hat{k}\cdot \vec{x}_1 v$ & & 13 & & $L_1+L_2+L_3+\hat{k}\cdot \vec{x}_1v+\hat{k}\cdot \vec{x}_2v$ & & 22 & & $L_2+2L_3+\hat{k}\cdot \vec{x}_2v+\hat{k}\cdot \vec{x}_3v$ \\
5 & & $L_1+L_2+2\hat{k}\cdot \vec{x}_2v$ & & 14 & & $L_1+\hat{k}\cdot \vec{x}_1v+\hat{k}\cdot \vec{x}_3v$ & & 23 & & $2L_1+L_2+2L_3+\hat{k}\cdot \vec{x}_2v+\hat{k}\cdot \vec{x}_3v$ \\
6 & & $L_1+L_2+2L_3+2\hat{k}\cdot \vec{x}_2v$ & & 15 & & $L_1+2L_2+\hat{k}\cdot \vec{x}_1v+\hat{k}\cdot \vec{x}_3v$ & & 24 & & $L_1+L_3+\hat{k}\cdot \vec{x}_2v+\hat{k}\cdot \vec{x}_3v$ \\
7 & & $L_3+2\hat{k}\cdot \vec{x}_2 v$ & & 16 & & $L_1+2L_3+\hat{k}\cdot \vec{x}_1v +\hat{k}\cdot \vec{x}_3v$ & & 25 & & $L_1+2L_2+L_3+\hat{k}\cdot \vec{x}_2v+\hat{k}\cdot \vec{x}_3v$ \\
8 & & $2L_1+2L_2+L_3+2\hat{k}\cdot \vec{x}_2v$ & & 17 & & $L_1+2L_2+2L_3+\hat{k}\cdot \vec{x}_1v+\hat{k}\cdot \vec{x}_3v$ & &   & &\\
9 & & $\hat{k}\cdot \vec{x}_1 v +\hat{k}\cdot \vec{x}_2 v$ & & 18 & & $L_2+L_3+\hat{k}\cdot \vec{x}_1v+\hat{k}\cdot \vec{x}_3v$ & &  & &\\
\hline
\end{tabular}
}
\end{table}

\section{Noise matrix}~\label{ap:noise}
The noise matrix of Sagnac combinations is given by
\begin{align}
    S_{\alpha} 
    & = \begin{pmatrix}
        S_{\alpha\alpha} & S_{\alpha\beta} & S_{\alpha\gamma} \\
        S_{\beta\alpha} & S_{\beta\beta} & S_{\beta\gamma} \\
        S_{\gamma\alpha} & S_{\gamma\beta} & S_{\gamma\gamma} 
    \end{pmatrix} .
\end{align}
where $S_{\alpha\alpha}$ denotes the auto-correlation PSD of the $\alpha$ combination, and 
$S_{\alpha\beta}$ the cross-correlation PSD between the $\alpha$ combination and $\beta$ combination, and so on.
Under the assumption of equal arm lengths, their expressions are given by~\cite{Hartwig:2023pft}
\begin{align}
    S_{\alpha\alpha}&=S_{\beta\beta}=S_{\gamma\gamma}=6S_{\text{oms}}+4\left[3-2\cos\delta-\cos(3\delta)\right]S_{\text{acc}},\\
    S_{\alpha\beta}&=S_{\alpha\gamma}=S_{\beta\gamma}=2[2\cos\delta+\cos(2\delta)]S_{\text{oms}}-4(1-\cos\delta)S_{\text{acc}},
\end{align}
where $\delta=2\pi fL$. 
Here, we assume that the noise is dominated by two components, the optical metrology system noise and test mass acceleration noise.
Their parameterized forms are given by~\cite{Babak:2021mhe}
\begin{align}
    S_{\text{oms}}&=\left(\frac{2\pi fs_{\text{oms}}}{c}\right)^2 \left[1+\left(\frac{2\times10^{-3}~\text{Hz}}{f}\right)^4\right]~\textrm{Hz}^{-1},\\
    S_{\text{acc}}&=\left(\frac{s_{\text{acc}}}{2\pi f c}\right)^2
    \left[1+\left(\frac{0.4\times 10^{-3}~\text{Hz}}{f}\right)^2\right]\left[1+\left(\frac{f}{8\times10^{-3}~\text{Hz}}\right)^4\right]~\textrm{Hz}^{-1}.
\end{align}
For Taiji, we adopt $s_{\text{oms}}=8\times10^{-12}~\text{m}$, $s_{\text{acc}}=3\times10^{-15}~\text{m}\cdot\text{s}^{-2}$, and $L=3\times10^6~\text{km}$.
\bibliography{ref}
\end{document}